\documentclass[prd,onecolum,preprintnumbers,showpacs,amsmath,amssymb,eqsecnum,nofootinbib]{revtex4}
\usepackage{graphicx}
\usepackage{dcolumn}
\usepackage{bm}
\preprint{DPNU-05-05}
\begin{document}
\title{Electromagnetic radiation due to naked singularity formation \\in self-similar
gravitational collapse}

\author{Eiji Mitsuda}
\email{emitsuda@gravity.phys.nagoya-u.ac.jp}

\author{Hirotaka Yoshino}
\email{yoshino@gravity.phys.nagoya-u.ac.jp}

\author{Akira Tomimatsu}
\email{atomi@gravity.phys.nagoya-u.ac.jp}
\affiliation{Department of Physics, Graduate School of Science, Nagoya
University, Chikusa, Nagoya 464-8602, Japan}
\begin{abstract}
Dynamical evolution of test fields in background geometry with a naked
 singularity is an important problem relevant to the Cauchy horizon
 instability and the observational signatures different from black hole
 formation. In this paper we study electromagnetic perturbations
 generated by a given current distribution in collapsing matter under a
 spherically symmetric self-similar background. Using the Green's
 function method, we construct the formula to evaluate the outgoing
 energy flux observed at the future null infinity. The contributions
 from ``quasi-normal'' modes of the self-similar system as well as
 ``high-frequency'' waves are clarified. We find a characteristic power-law time evolution of the outgoing energy flux which appears just before naked singularity formation, and give the criteria as to whether or not the outgoing energy flux diverges at the future Cauchy horizon. 
\end{abstract}
\maketitle
\section{Introduction\label{sec:intro}}
While Penrose has proposed the so-called cosmic censorship conjecture\cite{PenroseR:RNC1:1969}, several spacetimes admitting naked singularity formation even in gravitational collapse of physically reasonable matter from regular initial data have been found over the past three decades (see \cite{HaradaT:P53:1999, HaradaT:PTP107:2002} for a recent review). The examples well-studied are spherically symmetric collapses of an inhomogeneous dust ball\cite{SinghTP:CQG13:1996, JhinganS:CQG13:1996}, and a self-similar isothermal gas\cite{AOri:PRD42:1990, FoglizzoT:PRD48:1993, CarrBJ:PRD67:2003,CarrBJ:CQG18:2001}. A shell-focusing naked singularity can appear at the center in a wide range of the initial data set. 

It is an important problem to understand peculiar physical phenomena associated with naked singularity formation. Time evolution of various perturbations in black hole geometry have been extensively studied, and the late-time behaviors such as quasi-normal ringings and power-law tails have been clarified. Such perturbation analysis in background geometry involving a naked singularity will be interesting not only in relation to the possible instability of the Cauchy horizon but also in terms of revealing some typical patterns of time evolution observable as a precursor of naked singularity formation. 

The first step to approach this problem will be to consider spherically symmetric self-similar models of gravitational collapse as background spacetime. Because all dimensionless components of the background metric depend only on the one self-similar variable $z\equiv r/t$, the perturbation analysis becomes mathematically simpler. In addition, there exist numerical simulations\cite{HaradaT:PRD63:2001} showing that the geometrical structure and the fluid motion at late stages in general non-self-similar collapse of an isothermal gas can be well described by the general relativistic version\cite{AOri:PRD42:1990} of the Larson-Penston self-similar solution. We can expect self-similar solutions to be a realistic model of gravitational collapse.

Several works have been devoted to the analysis of perturbations
generated in spherically symmetric self-similar collapse with naked
singularity formation. For example, using quantum theory of particle
production in curved spacetime, it has been shown that the energy flux
of the semiclassical radiation diverges on the Cauchy horizon according
to inverse square power-law of the retarded
time\cite{HiscockWA:PRD26:1982, BarveS:NPB532:1998, BarveS:PRD58:1998,
VazC:PLB442:1998, SinghTP:PLB481:2000.2, MiyamotoU:PRD69:2004}. On the
other hand, considering classical perturbations of a massless scalar
field, Nolan and Waters\cite{NolanBC:PRD66:2002} have claimed that the
energy flux of scalar perturbations remains finite even at the Cauchy
horizon. Their analysis is based on the behavior of perturbations
written by $v^{n}H_{n}(z)$ with the advanced null coordinate $v$ and a
complex parameter $n$. In this paper we pursue the classical analysis
more extensively, by using the Green's function method. Though we use a
standard Fourier decomposition of the Green's function by the function
$\exp(i\omega\log|t|)$ (in an analogous way to the Mellin transformation
in \cite{NolanBC:PRD66:2002}), we reveal time evolution of perturbations by integrating the Fourier components with respect to the spectral parameter $\omega$, which will allow us to find new features missed in \cite{NolanBC:PRD66:2002}.

Further we focus our investigation on electromagnetic perturbations
generated by any given axisymmetric and circular current distribution in
collapsing matter. Of course we must assume that the existence of the
source current does not disturb the background self-similar metric. This
special choice of a test field is partly motivated by an astrophysical
interest related to highly energetic phenomena such as $\gamma$-ray
bursts (for example, see \cite{CunninghamCT:AJ224:1978} for
electromagnetic radiation in a process of black hole formation). In addition, the mathematical setup of the Green's function method  for electromagnetic perturbations becomes quite simpler in comparison with gravitational perturbations. Nevertheless, we would like to mention that the Green's function method developed in the following sections does not essentially rely on a special property of electromagnetic fields. The application to scalar and gravitational perturbations will be straightforward.

In this paper we are mainly concerned with time variation of the outgoing energy flux (namely, the Poynting flux) observed at the future null infinity. In Sec.~\ref{sec:setup}, we illustrate spherically symmetric self-similar models admitting naked singularity formation at the center, and we derive the basic equation for a gauge invariant and odd parity electromagnetic perturbation. In Sec.~\ref{sec:Green}, we introduce the retarded Green's  function and its Fourier decomposition, from which we construct the formula to extract the outgoing wave part of the electromagnetic perturbation propagating to the future null infinity. Contributions from various Fourier components parameterized by $\omega$, (for example, corresponding to high-$\omega$ waves, quasi-normal modes of the self-similar system with a complex $\omega$) are clarified in Sec.~\ref{sec:decomposition}. A characteristic power-law behavior (accompanied with or without an oscillatory behavior) of the outgoing energy flux as a function of the retarded time is found as a signature of naked singularity formation. We obtain in the final section the critical conditions for the self-similar background as to whether or not the outgoing energy flux diverges at the future Cauchy horizon, and a physical interpretation of our results is presented. Throughout this paper, the units in which $c=G=1$ are used.

\section{Setup of the system \label{sec:setup}}
Let us begin with a brief review of spherically symmetric self-similar
collapse with naked singularity formation. The line element of the
background spacetime is given by
\begin{equation}
 ds^{2}=-e^{2\nu}dt^{2}+e^{2\lambda}dr^{2}+r^{2}S^{2}(d\theta^{2}
+\sin^{2}{\theta}d\phi^{2}),
\label{eq:metric1}
\end{equation}
with the comoving coordinates $t$ and $r$. The self-similarity
considered here means
\begin{equation}
 \nu=\nu(z), \qquad \lambda=\lambda(z),\qquad  S=S(z),\label{eq:self-similarity}
\end{equation}
where $z\equiv r/t$, and the Ricci scalar has the form
\begin{equation}
 R =\frac{1}{t^{2}} \times (\text{function depending only on
  }z),
\end{equation}
which should grow up as $t$ approaches zero along a $z=$ constant line. Thus a singularity will appear at the center $r=0$ at the time $t=0$. A spacelike singularity may also appear along the $z=z_{*}$ line where $S(z_{*})=0$.

\begin{figure}
\includegraphics[height=7cm]{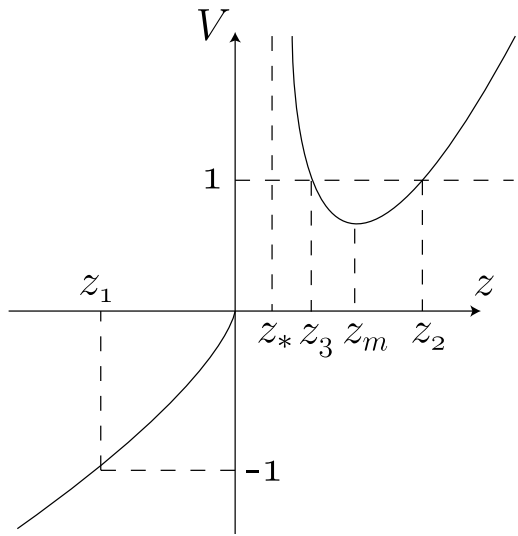}
\caption{Schematic description of the function $V(z)$ giving a typical
 collapse with naked singularity formation. In this figure it should be
 noted that $V(0)=0$ and $V\rightarrow\pm\infty$ as
 $z\rightarrow\pm\infty$.}
\label{fg:V}
\end{figure}

\begin{figure}
\includegraphics[height=7cm]{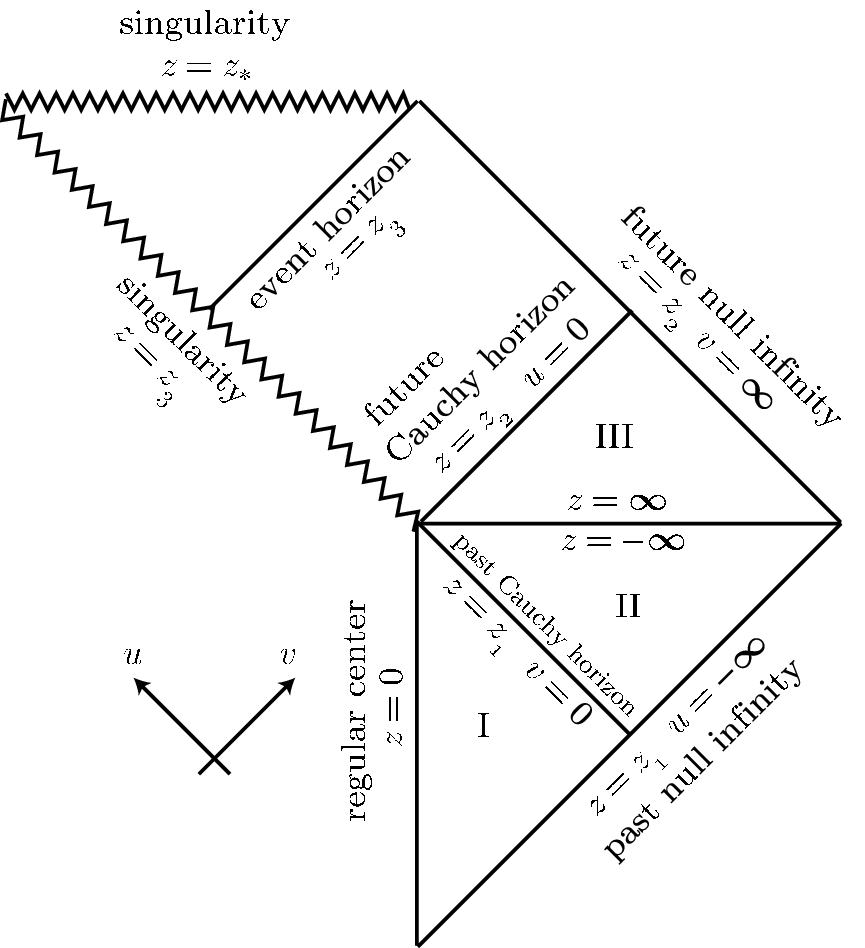}
\caption{The Penrose diagram of spacetime corresponding to the function $V(z)$ shown in Fig.~\ref{fg:V}. The values of the null coordinates $u$ and $v$ are given at some boundaries of the regions I, II and III.}
\label{fg:diagram}
\end{figure}

Using the function defined by
\begin{equation}
 V(z) \equiv ze^{\lambda-\nu},
\end{equation}
the equations for the radial outgoing and ingoing null geodesics are given by
\begin{equation}
 \frac{dz}{dt}=\pm\frac{z(1\pm V)}{tV}\label{eq:geodesics}.
\end{equation}
The key point is that the $z=$ constant line giving $V=1$ (or $V=-1$)
becomes a radial outgoing (or ingoing) null geodesic. In order to allow
the central singularity to be naked as a result of collapse with regular
initial data, in this paper, we assume the schematic behavior of the
function $V(z)$ drawn in Fig.~\ref{fg:V}, 
for which we obtain the global structure
of the self-similar background written in Fig.~\ref{fg:diagram} (see \cite{CarrBJ:PRD67:2003} for details of
how to construct kinematically the spacetime diagram). 

The value of local minimum of $V$ at $z=z_{m}$ should be less than
unity, because the line $z=z_{2}>z_{m}$ giving $V=1$ becomes the future
Cauchy horizon for the null naked singularity corresponding to another
root $z=z_{3}<z_{m}$ of $V=1$. There exist the future null infinity and
the event horizon also at the $z=z_{2}$ and $z=z_{3}$ null lines,
respectively. The equation $V=-1$ has a unique root $z=z_{1}<0$
corresponding to the past Cauchy horizon and the past null
infinity. Such a behavior of $V$ as a function of $z$ is found in a wide
parameter range of the self-similar metrics describing the collapse of
dust or isothermal gas
\cite{AOri:PRD42:1990, CarrBJ:CQG16:1999, BJCarr:PRD62:2000, CarrBJ:PRD62:2000}.

We are interested in time evolution of electromagnetic perturbations in
the regions I, II and III shown in Fig.~\ref{fg:diagram}. For this
purpose it is useful to introduce the retarded and advanced null
coordinates $u$ and $v$. Integrating Eq.~(\ref{eq:geodesics}), we obtain
\begin{equation}
 u= -|t|e^{h_{\text{out}}(z)}, \qquad v=\mp|t|e^{h_{\text{in}}(z)},
\end{equation}
where $v\leq 0$ in the region I and $v\geq 0$ in the region II and III,
while $u\leq 0$ in these regions. The functions $h_{\text{out}}(z)$ and
$h_{\text{in}}(z)$ are expressed by
\begin{equation}
 h_{\text{out}}(z)=\int \frac{Vdz}{z(V-1)}, \qquad  h_{\text{in}}(z)=\int \frac{Vdz}{z(V+1)},\label{eq:h}
\end{equation}
with the requirement that $h_{\text{out}}=h_{\text{in}}=0$ at $z=0$
(namely, $u=v=t<0$ at the regular center $r=0$ represented by the $z=0$
line). We can also note that the null coordinates are given by $u=v=r$
at the spacelike hypersurface $t=0$ corresponding to the horizontal
$z=\pm\infty$ line, which is the boundary between the regions II and
III. In the limit $z\rightarrow z_{1}$ the function $h_{\text{in}}$ is
approximately given by
\begin{equation}
 h_{\text{in}}(z)\simeq-\frac{1}{z_{1}V'(z_{1})}\log|z-z_{1}|.\label{eq:hin}
\end{equation}
Hereafter a prime means the derivative with respect to $z$. 
Because $-z_{1}V'(z_{1})$ is positive (see Fig.~\ref{fg:V}), the advanced null coordinate $v$ becomes
continuously zero, even if not analytically, at the boundary $z=z_{1}$
(namely, the past Cauchy horizon) between the regions I and II. It
remains finite at the past null infinity where $z=z_{1}$ and $t\sim
u=-\infty$. On the other hand, in the limit $z\rightarrow z_{2}$ we
obtain
\begin{equation}
 h_{\text{out}}(z)\simeq\frac{1}{z_{2}V'(z_{2})}\log|z-z_{2}|.\label{eq:hout}
\end{equation}
Because $z_{2}V'(z_{2})$ is also positive, the retarded null coordinate
$u$ also becomes
continuously zero at the boundary $z=z_{2}$ (namely, the future Cauchy
horizon) of the regions III, though it remains finite at the future null
infinity where $z=z_{2}$ and $t\sim v=\infty$. It is easy to check that
in the limit $\nu=\lambda=0$, the coordinates $u$ and $v$ become
identical with the usual null coordinates giving the Minkowski metric. 

The values of $-z_{1}V'(z_{1})$ and $z_{2}V'(z_{2})$ in
Eqs.~(\ref{eq:hin}) and (\ref{eq:hout}) will become
important to determine the behavior of the outgoing energy flux of the
perturbations near the future Cauchy horizon.
In the spherically symmetric spacetime written by the self-similar metric
(\ref{eq:self-similarity}), 
there exists a homothetic Killing vector field $\chi^{a}=(-t, -r, 0, 0)$, 
which satisfies
\begin{equation}
 \mathcal{L}_{\chi}g_{ab}=-2g_{ab},
\end{equation}
where $\mathcal{L}_{\chi}$ denotes the Lie derivative along the field
$\chi^{a}$ (the minus sign of the $t$ and $r$ components of $\chi^{a}$ is
chosen so that $\chi^{a}$ points toward the direction toward which the
curvature increases\cite{CarrBJ:PRD67:2003}).
The values of $-z_{1}V'(z_{1})$ and $z_{2}V'(z_{2})$ are derived by
evaluating the derivative of
the inner product of the homothetic Killing vector $\chi^{a}$ on the
past Cauchy horizon $z=z_{1}$ and the future Cauchy horizon $z=z_{2}$ as follows,
\begin{equation}
 \nabla^{a}(\chi^{b}\chi_{b})=-2\kappa\chi^{a},\label{eq:chi}
\end{equation} 
where $\kappa=-z_{1}V'(z_{1})$ on the past Cauchy horizon and
$\kappa=z_{2}V'(z_{2})$ on the future Cauchy horizon.
In the stationary black hole geometry, 
the constant $\kappa$ evaluated on the event horizon
for a Killing field $\chi^{a}$ normal to the horizon has a physical
meaning as the surface gravity.
Though $-z_{1}V'(z_{1})$ and $z_{2}V'(z_{2})$ do not necessarily have
such a physical meaning, in this paper we denote $-z_{1}V'(z_{1})$ and
$z_{2}V'(z_{2})$ by $\kappa_{1}$ and $\kappa_{2}$, respectively.

Finally in this section we derive the basic equation for electromagnetic
perturbations in the self-similar background, by considering the Maxwell
equation $\nabla^{b}(\partial_{b}A_{a}-\partial_{a}A_{b})=4\pi j_{a}$
for the vector potential $A_{a}$. The given source current $j_{a}$
satisfying the continuity equation $\nabla_{a}j^{a}=0$ is assumed to be axisymmetric and
circular, and in order to separate the variable $\theta$, we resolve the
source current into spherical harmonics. Then, the multipole component
parameterized by $l=1,2, \cdots$ is given by
\begin{equation}
  j_{a}=\left(0,\ 0,\ 0,\ j_{l} (t,r)\sin{\theta}\frac{\partial
	 P_{l}}{\partial \theta } \right)\label{eq:ja},
\end{equation}
which generates the vector potential $A_{a}$ (with the angular momentum
parameterized by $l$) of the form
\begin{equation}
 A_{a}=\left(0,\ 0,\ 0,\ a (t,r)\sin{\theta}\frac{\partial
	P_{l}}{\partial \theta } \right),\label{eq:Aa}
\end{equation} 
according to the equation
\begin{equation}
 \hat{L}a(t,z)=-Q_{l}(t,z),\label{eq:wave}
\end{equation}
where
\begin{eqnarray}
&&\hat{L}= t^{2}\frac{\partial^{2} }{\partial t^{2}}
-2zt\frac{\partial^{2} }{\partial t\partial z}
-\left(e^{2(\nu-\lambda )}-z^{2}\right)\frac{\partial^{2} }{\partial z^{2}}
+\left\{ 2z+z^{2}(\lambda^{'}-\nu^{'})-e^{2(\nu-\lambda )}(\nu^{'}-\lambda^{'})\right\} \frac{\partial }{\partial z} \nonumber \\
     &&\hspace{8cm}-(\lambda^{'}-\nu^{'})zt\frac{\partial }{\partial
      t}+\frac{e^{2\nu}l(l+1)}{z^{2}S^{2}},
\end{eqnarray}
and
\begin{equation}
 Q_{l}(t,z) = 4\pi t^{2}e^{2\nu}j_{l}.\label{eq:Ql}
\end{equation}
The current distribution $j_{l}(t,r)$ in Eq.~(\ref{eq:ja}) is not
specified in this paper, except that it rapidly decreases to zero at the
regular center $r=0$, at spatial infinity $r\rightarrow\infty$ and at
distant past $t\rightarrow -\infty$. We will obtain the solution $a(t,z)$ under the
boundary condition such that $a(t,z)$ remains regular at the regular
center $z=0$ (and $t<0$), and there is no ingoing flux from the past
null infinity $z=z_{1}$ (and $t\rightarrow-\infty$). The latter is
required to assure us that electromagnetic perturbations are generated
through the process of the self-similar gravitational collapse.

\section{The Green's function method \label{sec:Green}}
As mentioned in Sec.~\ref{sec:intro}, the main purpose of this paper is to estimate the outgoing  energy flux (the Poynting flux) observed at the future null infinity near the future Cauchy horizon, by solving the wave equation (\ref{eq:wave}) for the electromagnetic vector potential $a(t,z)$. Our approach to this problem is to develop the Green's function method useful for the analysis of any test fields in the self-similar background. 

Following the standard technique, we give the time evolution of $a(t,z)$ as
\begin{equation}
 a(t,z)=\int\int G(t,z|s,y)Q_{l}(s,y)dsdy,\label{eq:a}
\end{equation}
with the (time-domain) Green's function $G(t,z|s,y)$, satisfying
\begin{eqnarray}
\hat{L}G(t,z|s,y)=-\delta(t-s)\delta(z-y).
      \label{eq:Maxwell3}
\end{eqnarray}
The causality condition requires that $G=0$ if the source point
represented by the coordinates $s$ and $y$ is located in the exterior of
the past light cone of the observation point represented by $t$ and $z$,
in other words, if $u'>u$ or $v'>v$ for the null coordinates defined by
$u'\equiv u(s,y)$ and $v' \equiv v(s,y)$. 

The key point for analyzing the Green's function is that the
differential operator $\hat{L}$ in Eq.~(\ref{eq:Maxwell3}) is invariant
under the scale transformation $t\rightarrow kt$ with a parameter $k$,
for which $z$ is fixed. This motivates us to adopt the standard Fourier
decomposition of the form
\begin{equation}
 G(t,z|s,y)=\frac{1}{2\pi s}\int^{\infty}_{-\infty}\tilde{G}(z,y,\omega)e^{i\omega\log{|t/s|}}d\omega,\label{eq:G}
\end{equation}
because Eq.~(\ref{eq:Maxwell3}) reduces to the tractable ordinary
differential equation
\begin{equation}
 \tilde{G}''(z,y,\omega)+p(z,\omega)\tilde{G}'(z,y,\omega)+q(z,\omega)\tilde{G}(z,y,\omega)=\frac{\delta(z-y)}{e^{2(\nu-\lambda)}(1-V^{2})},\label{eq:ODE}
\end{equation}
where the coefficients are given by
\begin{equation}
 p(z,\omega)= \frac{2(i\omega-1)z+(\nu^{'}-\lambda^{'})e^{2(\nu-\lambda)}(1+V^{2})}{e^{2(\nu-\lambda)}(1-V^{2})},\label{eq:p}
\end{equation}
and
\begin{equation}
 q(z,\omega)= \frac{\omega(i+\omega)S^{2}z^{2}-i\omega(\nu^{'}-\lambda^{'})S^{2}z^{3}-e^{2\nu}l(l+1)}{S^{2}z^{2}e^{2(\nu-\lambda)}(1-V^{2})}
  .\label{eq:q}
\end{equation}
The spectral parameter $\omega$ introduced here may be regarded as a wave frequency  in the lapse of the logarithmic time $\log|t|$. Hence, hereafter we will call $\tilde{G}(z,y,\omega)$ the frequency-domain Green's function as usual.

It is straightforward to construct the frequency-domain Green's function
by the help of  two independent homogeneous solutions for
Eq.~(\ref{eq:ODE}). If the boundary condition mentioned in the previous
section is imposed on $a(t,z)$, one of the homogeneous solution denoted
by $\psi_{0}(z,\omega)$ should be regular at the regular center $z=0^{-}$,
while the other denoted by $\psi_{\text{out}}(z,\omega)$ should be purely
outgoing at $z=z_{1}<0$ to assure the absence of ingoing waves
originated from the past null infinity. Then, we have
\begin{eqnarray}
 \tilde{G}(z,y,\omega)= \left \{ 
\begin{array} {ll}
\tilde{G}_{1}(z,y,\omega)=\psi_{\text{out}}(z,\omega)\psi_{0}(y,\omega)/\bar{W}(y,\omega)
 & \mbox{for $z<y$}, \\
\tilde{G}_{2}(z,y,\omega)=\psi_{0}(z,\omega)\psi_{\text{out}}(y,\omega)/\bar{W}(y,\omega) & \mbox{for $z>y$},
\end{array} \right.\label{eq:tildeG}
\end{eqnarray}
where the common factor
$\bar{W}(y,\omega)=w(\omega)e^{\nu(y)-\lambda(y)}e^{i\omega
h_{\text{in}}(y)}e^{i\omega h_{\text{out}}(y)}$ is derived from the
Wronskian
\begin{equation}
 \psi_{\text{out}}\psi'_{0}-\psi'_{\text{out}}\psi_{0}=w(\omega)\exp\left(-\int
		 p(z',\omega)dz'\right)=\frac{\bar{W}}{e^{2(\nu-\lambda)}(1-V^{2})},
\label{eq:W1}
\end{equation}
with $w(\omega)$ independent of $z$.

If a homogeneous  solution which becomes purely ingoing at $z=z_{1}$ is
denoted by $\psi_{\text{in}}(z,\omega)$, the mode $\psi_{0}$ regular at
$z=0^{-}$ may be written by the sum
\begin{equation}
 \psi_{0}(z,\omega)=  \psi_{\text{out}}(z,\omega) +\psi_{\text{in}}(z,\omega).
\end{equation}
In this paper we claim the wave modes $\psi_{\text{out}}$ and
$\psi_{\text{in}}$ to become purely outgoing and ingoing, respectively, at
$z=z_{1}$ in the sense that  they can be expressed by the WKB forms
\begin{equation}
 \psi_{\text{out}}(z,\omega)=g_{\text{out}}(z,\omega)\exp\left\{i\omega h_{\text{out}}(z)\right\}, \qquad  \psi_{\text{in}}(z,\omega)=g_{\text{in}}(z,\omega)\exp\left\{i\omega h_{\text{in}}(z)\right\},
\end{equation}
with the amplitudes $g_{\text{out}}$ and $g_{\text{in}}$ ``regular'' at
$z=z_{1}$ (where $V=-1$ and $h_{\text{in}}\rightarrow
-\infty$). However, in general, we cannot expect that $g_{\text{out}}$
and  $g_{\text{in}}$  remain regular at another regular singular point
$z=z_{2}$ (where $V=1$ and $h_{\text{out}}\rightarrow -\infty$) in
Eq.~(\ref{eq:ODE}). For example, $g_{\text{out}}$ may contain a term
with the oscillatory factor $\exp(-i\omega h_{\text{out}} )$ which
becomes singular in the limit $z\rightarrow z_{2}$. This corresponds to
a partial conversion of outgoing waves into ingoing ones in the
propagation from the past Cauchy horizon $z=z_{1}$ toward the future
null infinity $z=z_{2}$, and may be interpreted as a result of back
scattering effect of spacetime curvature. (In the later discussion we
will take account of this mode conversion to estimate electromagnetic
radiation observed at the future null infinity.) Then, denoting two
independent modes which become purely outgoing and ingoing at $z=z_{2}$
by $\bar{\psi}_{\text{out}}$  and $\bar{\psi}_{\text{in}}$, respectively, we obtain 
\begin{equation}
 \psi_{\text{out}}(z,\omega)=b_{1}(\omega)\bar{\psi}_{\text{out}}(z,\omega)+
b_{2}(\omega)\bar{\psi}_{\text{in}}(z,\omega),\label{eq:fout}
\end{equation}
with some coefficients $b_{1}$ and $b_{2}$ dependent on $\omega$. These
new modes $\bar{\psi}_{\text{out}}$  and $\bar{\psi}_{\text{in}}$ also can be
written as
\begin{equation}
 \bar{\psi}_{\text{out}} = \bar{g}_{\text{out}}(z,\omega)\exp\left\{i\omega h_{\text{out}}(z)\right\},
  \qquad  \bar{\psi}_{\text{in}}=\bar{g}_{\text{in}}(z,\omega)\exp\left\{i\omega h_{\text{in}}(z)\right\},\label{eq:barf}
\end{equation}            
with the amplitudes $\bar{g}_{\text{out}}$  and $\bar{g}_{\text{in}}$
regular at $z=z_{2}$, and the relation 
\begin{equation}
 \psi_{0}(z, \omega) = \bar{\psi}_{\text{out}}(z, \omega)+\bar{\psi}_{\text{in}}(z, \omega),\label{eq:f02}
\end{equation}
is also assumed for their normalization. 

Now let us present a more definite expression of the double integral in
Eq.~(\ref{eq:a}) over the regions I, II and III shown in Fig.~\ref{fg:diagram}. Recall
that the frequency-domain Green's function $\tilde{G}(z,y,\omega)$ is
constructed under the conditions at the inner boundary $z=0$ and the
outer one $z=z_{1}$ surrounding the region I. Then, if the observation
point $(t, z)$ is located in the region I, the solution $a(t,z)$ can be
written as
\begin{equation}
 a(t,z)=\int_{0}^{z}dy\int_{-\infty}^{s_{1}}ds \ G_{1}(t,z|s,y)Q_{l}(s,y)+\int_{z}^{z_{1}}dy\int_{-\infty}^{s_{2}}ds \ G_{2}(t,z|s,y)Q_{l}(s,y),\label{eq:a2}
\end{equation}
where $G_{1}(t,z|s,y)$ and $G_{2}(t,z|s,y)$ are given by the
Fourier-type integral (\ref{eq:G}) of the frequency-domain Green's
functions $\tilde{G}_{1}(z,y,\omega)$ and $\tilde{G}_{2}(z,y,\omega)$,
respectively. From the causality condition the upper limits $s_{1}(u,y)$
and $s_{2}(v,y)$ in the integrals with respect to $s$ should be
determined by the relations $u(t,z)=u'(s_{1},y)$ and
$v(t,z)=v'(s_{2},y)$. 
Note that the integral of the second term of the right hand side with
respect to $s$ approaches the integral in the range from $s=-\infty$ to
$s=t$ as $z\rightarrow z_{1}$ (and $v\rightarrow 0$).

The integral representation (\ref{eq:a2}) assures the absence of
unphysical ingoing radiation which may make the solution $a(t,z)$
singular at the past Cauchy horizon $z=z_{1}$ (or $v=0$), where the
second term (including $G_{2}$) vanishes, and owing to the relation
$\tilde{G}_{1} \sim \psi_{\text{out}}(z,\omega)$ the first term (including
$G_{1}$) becomes regular. This regularity of $a(t,z)$ at the outer
boundary $z=z_{1}$ of the region I allows us to extrapolate
Eq.~(\ref{eq:a2}) to the form valid at the observation point $(t,z)$
located in the region II where $-\infty\leq z\leq z_{1}$ and $t\leq 0$
as follows,
\begin{eqnarray}
 a(t,z)&=&\int_{0}^{z_{1}}dy\int_{-\infty}^{s_{1}}ds~G_{1}(t,z|s,y)Q_{l}(s,y)\nonumber\\
&&+\int_{z_{1}}^{z}dy\int_{s_{2}}^{s_{1}}ds~\left\{G_{1}(t,z|s,y)-
 G_{2}(t,z|s,y)\right\}Q_{l}(s,y).
\label{eq:a3}
\end{eqnarray}
The first term in Eq.~(\ref{eq:a3}) corresponds to the contribution from
the source point $(s,y)$ located in the region I, while the second one
corresponds to the contribution from the source point located in the
region II. Note that the integral of the second term with respect to $s$
is limited to the range $s_{2}\leq s \leq s_{1}$ by virtue of the
causality condition, 
and approaches the integral in the range from $s=-\infty$ to $s= t$ as
$z\rightarrow z_{1}$ (and $v\rightarrow 0$) to allow us to continuously
shift from the integral representation (\ref{eq:a2}) to that of
Eq.~(\ref{eq:a3}) beyond $v=0$.

The next step is to extrapolate Eq.~(\ref{eq:a3}) to the form valid in
the observation point $(t,z)$ located in the region III where
$+\infty\geq z\geq z_{2}$ and $t\geq 0$. As shown in
Fig.~\ref{fg:diagram}, the boundary between the regions II and III is
given by the $z=\pm\infty$ line. This change of the sign of the
variables $z$ (and $t$) is not troublesome for the continuous
extrapolation of Eq.~(\ref{eq:a3}), because we obtain
\begin{equation}
 \psi_{\text{out}}(z, \omega)\exp(i\omega\log|t|) \sim \psi_{\text{in}}(z, \omega)\exp(i\omega\log|t|) \sim \exp(i\omega r),
\end{equation}
in the limit $z\rightarrow\pm\infty$ where we use $V\rightarrow\pm\infty$ in Eq.~(\ref{eq:h}). (The amplitudes $g_{\text{out}}(z,\omega)$ and $g_{\text{in}}(z,\omega)$ are also continuous at the boundary $z=\pm\infty$.) Hence, Eq.~(\ref{eq:a3}) holds in the region III without any modifications, except that the integral in the second term of the right-hand side should be understood as the contribution from the source point $(s,y)$ located in the regions II and III.

To clarify the physical properties of the solution $a(t,z)$, it will be
convenient to use the null coordinates $u$ and $v$ for the observation
point and $u'$ and $v'$ for the source point. This coordinate
transformation gives the Jacobian
\begin{equation}
 \left|\frac{\partial(s,y)}{\partial(u',v')}\right| \equiv J(u',v')=\frac{sy\left(1-V^{2}(y)\right)}{2V(y)u'v'},
\end{equation} 
and Eqs.~(\ref{eq:a2}) and (\ref{eq:a3}) are rewritten into the unified
form 
\begin{equation}
 a(u,v)=\int_{-\infty}^{u}du' \int_{u'}^{v}dv' H(u,v|u',v') Q_{l}(u',v')J(u',v'),\label{eq:a4}
\end{equation}
where the lower limit of the integral with respect to $v'$ should be
$u'$, because the regular center $y=0$ corresponds to $v'=u'$. If the
observation point $(u,v)$ is located in the region I, the function
$H(u,v|u',v')$ in Eq.~(\ref{eq:a4}) is equal to the original Green's
function $G(t,z|s,y)$. However, if the observation point $(u,v)$ is
located in the region II (or in the region III), we obtain
\begin{equation}
 H(u,v|u',v') = G_{1}(t,z|s,y),\label{eq:H1}
\end{equation}
for the source point $(u',v')$ located in the region I, and
\begin{equation}
 H(u,v|u',v') = G_{1}(t,z|s,y)-G_{2}(t,z|s,y),\label{eq:H2}
\end{equation}
for the source point $(u',v')$ located in the region II (or in the
regions II and III). The validity of the integral representation (\ref{eq:a4}) is supported in Appendix~\ref{sec:Minkowski}, by applying this formula to a static field $a(t,z)=a(r)$ in the Minkowski background.

Our main concern in this paper is the emission of outgoing radiation
toward the future null infinity. Hence, using the two modes
$\bar{\psi}_{\text{out}}(z, \omega)$ and $\bar{\psi}_{\text{in}}(z,\omega)$
instead of $\psi_{\text{out}}(z,\omega)$ and $\psi_{\text{in}}(z,\omega)$, we
divide the frequency-domain Green's function given by
Eq.~(\ref{eq:tildeG}) into the outgoing and ingoing wave parts as
follows,
\begin{equation}
 \tilde{G}_{\gamma}(z,y,\omega)=\bar{\psi}_{\text{out}}(z,\omega)\tilde{G}_{\gamma \text{out}}(y,\omega)+
\bar{\psi}_{\text{in}}(z,\omega)\tilde{G}_{\gamma \text{in}}(y,\omega),\label{eq:tildeGgamma}
\end{equation}
where $\gamma=1, 2$. The outgoing part of the right-hand side of
Eq.~(\ref{eq:tildeGgamma}) can derive the ``retarded'' Green's functions
$G_{\gamma \text{out}}(u,v|u',v')$ as
\begin{equation}
 G_{\gamma \text{out}}(u,v|u',v')= \frac{1}{2\pi
  s}\int^{\infty}_{-\infty}\bar{\psi}_{\text{out}}(z,\omega)\tilde{G}_{\gamma
  \text{out}}(y,\omega)e^{i\omega\log{(t/|s|)}}d\omega,
\end{equation}
which reduces to 
\begin{equation}
 G_{\gamma \text{out}}(u|u',v')=\frac{1}{2\pi
  s}\int^{\infty}_{-\infty}\bar{g}_{\text{out}}(z_{2},\omega)\tilde{G}_{\gamma
  \text{out}}(y,\omega)e^{i\omega\log{(-u/|s|)}}d\omega,
\end{equation}
in the limit $v\rightarrow\infty$ (i.e., $z\rightarrow z_{2}$) with a
fixed $u$. Via the above-mentioned procedure and Eq.~(\ref{eq:a4}) we
can obtain the outgoing part $a_{\text{out}}(u,v)$ of the vector
potential $a(u,v)$. The final formula for the outgoing radiation field
$a_{\text{out}}(u)\equiv a_{\text{out}}(u,z_{2})$ observed at the future
null infinity $v\rightarrow\infty$ (i.e., $z\rightarrow z_{2}$) is given
by
\begin{equation}
 a_{\text{out}}(u)=\int^{u}_{-\infty}du'\int^{\infty}_{u'}dv'H_{\text{out}}(u|u',v')Q_{l}(u',v')J(u',v'),
\end{equation}
where $H_{\text{out}}(u|u',v')$ is equal to $G_{1 \text{out}}(u|u',v')$
for the source point $(u',v')$ located in the region I, and is equal to
$G_{1 \text{out}}(u|u',v')-G_{2 \text{out}}(u|u',v')$ for the source
point $(u',v')$ located in the regions II and III. 
The explicit form may be written by
\begin{equation}
 H_{\text{out}}(u|u',v')=\frac{1}{2\pi se^{\nu(y)-\lambda(y)}}\int_{-\infty}^{\infty}\frac{\bar{g}_{\text{out}}(z_{2},\omega)}{w(\omega)}\left\{K(y,\omega)e^{i\omega\log(-u/|v'|)}+
N(y,\omega)e^{i\omega\log(u/u')}\right\}d\omega,\label{eq:Hout}
\end{equation}
where 
\begin{equation}
 K=0, \qquad N=(b_{1}-b_{2})\bar{g}_{\text{in}}(y,\omega),
\end{equation}
in the range $v' \geq -u'$ (i.e., in the region III). If the source
point is located in the region I or II, the functions $K$ and $N$ should
be expressed by $g_{\text{out}}(y,\omega)$ and $g_{\text{in}}(y,\omega)$ instead of
$\bar{g}_{\text{out}}(y, \omega)$ and $\bar{g}_{\text{in}}(y,\omega)$ to
be regular at the boundary $y=z_{1}$ (i.e., $v'=0$). Hence, we obtain
\begin{equation}
 K=(b_{1}-1)g_{\text{out}}(y,\omega), \qquad N=b_{1}g_{\text{in}}(y,\omega),
\end{equation}
in the range $0\leq v' \leq -u'$ (i.e., in the region II), and
\begin{equation}
 K=b_{1}g_{\text{out}}(y,\omega), \qquad N=b_{1}g_{\text{in}}(y,\omega),
\end{equation}
in the range $u'\leq v' \leq 0$ (i.e., in the region I). It should be also remarked that from the causality condition we can use Eq.~(\ref{eq:Hout}) only in the range $u'\leq u$ for the retarded time $u'$ of the source point.

Because the physical quantity to be observable is the energy flux of the
outgoing radiation, we must calculate the derivative of $a(u)$ with
respect to the retarded time $u$. Here, for later convenience,  we
denote the derivative with respect to $\log(-u)$ by a dot, and we have
\begin{equation}
 \dot{a}_{\text{out}}(u)= \Psi(u)+\Phi(u),\label{eq:dotaout}
\end{equation}
where 
\begin{equation}
 \Psi(u)=\int^{u}_{-\infty}du'\int^{\infty}_{u'}dv'\dot{H}_{\text{out}}(u|u',v')Q_{l}(u',v')J(u',v'),\label{eq:A}
\end{equation}
and
\begin{equation}
 \Phi(u)=u\int^{\infty}_{u}dv'H_{\text{out}}(u|u,v')Q_{l}(u,v')J(u,v').\label{eq:B}
\end{equation}
The future Cauchy horizon due to naked singularity formation appears on
the boundary $z=z_{2}$ (i.e., at the retarded time $u=0$) of the region
III. Hence, to understand clearly a distinctive effect of the naked
singularity, our task in the following sections is to investigate the
asymptotic behavior of $\dot{a}_{\text{out}}(u)$ in the limit
$u\rightarrow 0$ for clarifying whether or not a burst-type emission of
the infinite energy flux can occur in the self-similar collapse.

\section{Decomposition of contributions to the Green's function\label{sec:decomposition}}

Now let us try to obtain explicitly the asymptotic behavior of the
functions $\Psi(u)$ and $\Phi(u)$ in the limit $u\rightarrow 0$ according to
the formalism given in the previous section. The first step would be to
calculate the integral with respect to $\omega$ in Eq.~(\ref{eq:Hout}), by
considering the analytic structure of the Wronskian factor $w$ in
Eq.~(\ref{eq:W1}) as a function of $\omega$. Hence, following the analysis of
test fields in black hole background, we decompose the Green's function
into several distinct parts corresponding to contributions from
high-frequency waves with $|\omega|\rightarrow\infty$ and ``quasi-normal''
modes of the system with complex frequencies giving $w=0$.

\subsection{High-frequency contribution}
In the high-frequency limit $|\omega|\rightarrow\infty$ we can apply the
WKB approximation to obtain the homogeneous solutions for
Eq.~(\ref{eq:ODE}), and at the leading order the amplitudes
$g_{\text{out}}(z,\omega)$ and $g_{\text{in}}(z,\omega)$ are found to
become constants denoted here by $c_{\text{out}}$ and
$c_{\text{in}}$, respectively, without a mode mixing (i.e.,
$b_{1}\simeq 1$, $b_{2}\simeq 0$). Then, the Wronskian factor $w$ is
approximately given by 
\begin{equation}
 w(\omega)\simeq 2i\omega c_{\text{out}}c_{\text{in}}.
\end{equation}
It becomes easy to calculate the derivative of $H_{\text{out}}$ with
respect to $\log(-u)$ (denoted by $\dot{H}^{(F)}_{\text{out}}$) under this
approximation, and we have
\begin{equation}
 \dot{H}^{(F)}_{\text{out}}(u|u',v')=\frac{1}{2se^{\nu(y)-\lambda(y)}}\left\{\frac{c_{\text{out}}}{c_{\text{in}}}\delta\left(\log{\left(\frac{u}{v'}\right)}\right)+\delta\left(\log{\left(\frac{u}{u'}\right)}\right)\right\},\label{eq:dotHoutF}
\end{equation}
which is applied to the estimation of $\Psi(u)$ in
Eq.~(\ref{eq:A}). Denoting the high-frequency part of $\Psi$ by $\Psi_{F}$, we
arrive at the result such that $\Psi_{F}(u)=\Psi_{F1}(u)+\Psi_{F2}(u)$, where 
\begin{equation}
 \Psi_{F1}(u) = \frac{c_{\text{out}}
  u}{2c_{\text{in}}}\int_{-\infty}^{u}\bar{Q}_{l}(u',u)J(u',u)du',
  \qquad \Psi_{F2}(u) = \frac{u}{4}\int_{u}^{\infty}\bar{Q_{l}}(u,v')J(u,v')dv',\label{eq:AF}
\end{equation}
with the integrand factor $\bar{Q_{l}}$ defined by $\bar{Q_{l}} \equiv  Q_{l}/se^{\nu-\lambda}=4\pi se^{\nu+\lambda}j_{l}$ for a given current distribution $j_{l}$.

\begin{figure}
\includegraphics[height=7cm]{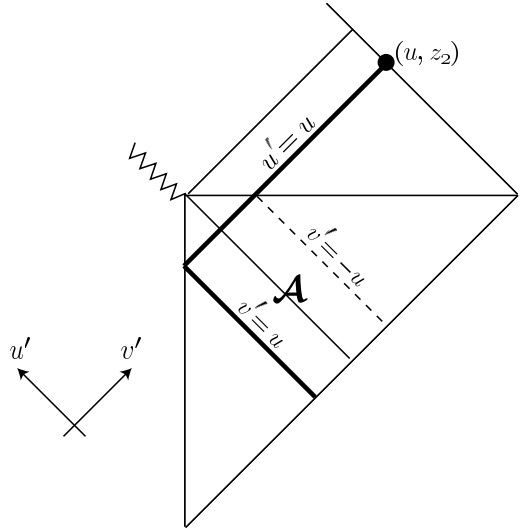}
\caption{The source region shown in the Penrose diagram, which is given by the null coordinates $u'$ and $v'$. The null geodesics $v'=u$ and $u'=u$ starting from the past null infinity and arriving at the observation point denoted by $(u,z_{2})$ are shown by the thick solid lines. The broken line corresponds to the $v'=-u$ line. The boundaries of the source region $\mathcal{A}$ are given by $u'=-\infty$, $v'=u$, $u'=u$ and $v'=-u$.}
\label{fg:idiagram}
\end{figure}

High-frequency waves will be able to propagate without scattering due to
the spacetime curvature and the centrifugal barrier (depending on
$l$). In fact, the $\delta$-functions which appear in Eq.~(\ref{eq:dotHoutF}) allow us
to use the approximation of geometrical optics for the propagation of
high-frequency waves to the future null infinity at the retarded time
$u$ (along the path drawn in Fig.~\ref{fg:idiagram}). 
The source points $(u',v')$ giving
the first term $\Psi_{F1}$ in $\Psi_{F}$ are located on the ingoing null line
$v'=u$. Hence, the outgoing flux represented by $\Psi_{F1}$ is a result of
the reflection (at the regular center $v'=u'=u$) of ingoing null rays
propagating along $v'=u$, which are generated by the  current
distribution $j_{l}$ at the source points $(u', u)$ ranging from $u'
\rightarrow -\infty$ to $u'=u$. On the other hand the second term
$\Psi_{F2}$ represents the outgoing flux which is generated by the current
distribution $j_{l}$ just on the past null cone $u'=u$ (ranging from $v'= u$ to $v' \rightarrow \infty$) and is directly propagating to the
observation point.
If the retarded time $u$ of observation at the future null infinity is
nearly equal to zero, the source points for $\Psi_{F1}$ (or $\Psi_{F2}$) on
the ingoing null line $v'=u$ (or the outgoing one $u'=u$) become nearly
identical with the past (or future) Cauchy horizon, where we obtain
roughly $u' \sim s$ (or $v' \sim s$) except an unimportant constant
factor. Then, the Jacobians $J$ in the integrands of Eq.~(\ref{eq:AF})
are roughly given by
\begin{equation}
 J(u',u) \sim \frac{1}{u}\left(\frac{u}{u'}\right)^{\kappa_{1}}, \qquad 
J(u,v') \sim \frac{1}{u}\left(\frac{-u}{v'}\right)^{\kappa_{2}},\label{eq:J2}
\end{equation}
where $\kappa_{1}$ and $\kappa_{2}$ are the values previously defined in Sec.~\ref{sec:setup}.
Hence we obtain
\begin{equation}
 \Psi_{F1}(u) \sim
  (-u)^{\kappa_{1}}\int_{-\infty}^{u}j_{l}(u',z_{1})(-u')^{1-\kappa_{1}}du',\qquad
 \Psi_{F2}(u) \sim
  (-u)^{\kappa_{2}}\int_{u}^{\infty}j_{l}(v',z_{2})v^{\prime 1-\kappa_{2}}dv'.\label{eq:AF2}
\end{equation} 
As was mentioned in Sec.~\ref{sec:setup}, we assume a rapid decrease of
the current distribution $j_{l}$ near the regular center. A plausible
choice would be that  $j_{l}(s,y) \simeq h(s)r^{2}(s,y)S^{2}(y)$ at the
source point $r=sy \simeq 0$, where $h(s)$ is introduced as some
amplification of $j_{l}$ in the process of the self-similar collapse. It
may be interesting to consider the case that $j_{l}(s,y)$ diverges at
the onset $s=0$ of the naked singularity for a fixed nonzero $y$
(namely, $hs^{2}\rightarrow\infty$ as $s\rightarrow 0$). However, our
purpose in this paper is to pursue the possibility of amplified
generation of electromagnetic radiation due to the self-similar dynamics
of background geometry involving a naked singularity, which requires us
to calculate $\Psi_{F1}$ and $\Psi_{F2}$ without any divergence of the source
current $j_{l}$. (Hereafter we assume $h$ to be finite at $s=0$ for
simplifying our discussion.) Furthermore, we obtain the relations
\begin{equation}
 \kappa_{1}=1-\frac{u^{2}R_{uu}}{8}, \qquad
  \kappa_{2}=1-\frac{v^{2}R_{vv}}{8},
\label{eq:energycondition}
\end{equation}
where $R_{vv}$ and $R_{uu}$ are the components of the Ricci tensor
evaluated at $z=z_{1}$ and $z=z_{2}$, respectively. Then, if the strong
energy condition holds for collapsing matter, we can require the
inequality
\begin{equation}
 \kappa_{1} \leq 1, \qquad \kappa_{2} \leq 1, 
\end{equation}
to show that the integrals in Eq.~(\ref{eq:AF2})
converge even in the limit $u\rightarrow 0$. (Of course, it should be
also assumed that the current distribution $j_{l}$ rapidly decreases as
the source point becomes close to the past or future null infinity
$u'\rightarrow -\infty$ or $v' \rightarrow \infty$). Hence, finally we
find the asymptotic power-law behaviors of $\Psi_{F1}$ and $\Psi_{F2}$ giving
the outgoing flux of high-frequency waves to be 
\begin{equation}
 \Psi_{F1}(u) \sim (-u)^{\kappa_{1}}, \qquad \Psi_{F2}(u) \sim (-u)^{\kappa_{2}}, 
\end{equation}
in the limit $u\rightarrow 0$.

Here let us turn our attention to the additional term $\Phi(u)$ in
Eq.~(\ref{eq:dotaout}). Note that the source points $(u',v')$ giving
$\Phi(u)$ are also located on the past null cone $u'=u$ in the same way as
$\Psi_{F2}$ in Eq.~(\ref{eq:AF}) for high-frequency waves. The contribution
to $\Phi(u)$ given at the retarded time $u$ will represent outgoing waves
generated on the past null cone ranging from $v' = u$ to $v' \rightarrow
\infty$. This allows us to estimate easily the asymptotic behavior of
$\Phi(u)$ in the limit $u\rightarrow 0$ without using the high-frequency
approximation. The key point is that in the limit $u\rightarrow 0$ the
range of the past null cone in the regions I and II shrinks to zero (see
Fig.~\ref{fg:idiagram}). Hence, using Eq.~(\ref{eq:Hout}) valid in the
region III, the function $H_{\text{out}}(u|u,v')$ in Eq.~(\ref{eq:B}) can be
given by
\begin{equation}
 H_{\text{out}}(u|u,v')=\frac{I}{2\pi se^{\nu(z_{2})-\lambda(z_{2})}},
  \qquad
 I = \int_{-\infty}^{\infty}\frac{(b_{1}-b_{2})}{w(\omega)}\bar{g}_{\text{out}}(z_{2},\omega)\bar{g}_{\text{in}}(z_{2},\omega) d\omega.\label{eq:Hout2}
\end{equation}
in the limit $u\rightarrow 0$, where $s \sim v'$ and $y=z_{2}$ for the
source point $u'=0$. The convergence of  the integral $I$ with
respect to $\omega$ in Eq.~(\ref{eq:Hout2}) may be subtle, because the
integrand is proportional to $1/\omega$ in the high-frequency
limit. However, the contribution of such a high-frequency part can be
canceled out, for example, if the infinite integral of any function
$\sigma(\omega)$ is defined as 
\begin{equation}
 \int_{-\infty}^{\infty} \sigma(\omega) d\omega \equiv \lim_{\omega_{0}\rightarrow\infty}\int^{\omega_{0}}_{-\omega_{0}} \sigma(\omega)d\omega.
\end{equation}  
By virtue of such a regularization of the integral we obtain
\begin{equation}
 \Phi(u)=\frac{2I}{\pi} \Psi_{F2}(u) \sim (-u)^{\kappa_{2}},\label{eq:B2}
\end{equation}
in the limit $u\rightarrow 0$. If wave frequencies are not so large,
outgoing waves will be efficiently back-scattered, and their
contribution to the function $H_{\text{out}}(u|u,v')$ in $\Phi(u)$ becomes
important to estimate the value of $I$. Nevertheless, from
Eq.~(\ref{eq:B2}) we can conclude that such a back-scattering effect is not
relevant to the asymptotic power-law behavior of $\Phi(u)$ which becomes
identical with that of $\Psi_{F2}(u)$.

\subsection{Contribution from poles in the Green's function}
The frequency-domain Green's function may contain singularities in the complex $\omega$-plane. In  black hole background it is well-known that late-time tails of massless test fields are generated by a low-frequency contribution to the frequency-domain Green's function, which appears as a $\omega=0$ branch cut in the Wronskian factor $w(\omega)$. There may be also a branch cut in the frequency-domain Green's function (\ref{eq:tildeG}) considered here. However, in this paper we do not pursue such a problem. Here our investigation is focused on the pole contribution to $\Psi(u)$, which will be interesting as a resonant behavior of test fields in the self-similar background. We use the residue calculus to evaluate the integral (with respect to $\omega$) in $\dot{H}_{\text{out}}$, from which the high-frequency part $\dot{H}^{(F)}_{\text{out}}$ is subtracted.

Recall that the simple poles in the frequency-domain Green's function
are given by the zeros of $w(\omega)$. We must remark that even in the
Minkowski background there are zeros at $\omega=\pm i$ (see
Appendix~\ref{sec:Minkowski}), and the two mode functions
$\psi_{\text{out}}(z,\pm i)$ and $\psi_{\text{in}}(z, \pm i)$ given at the
frequencies can satisfy the relation
$\psi_{0}=\psi_{\text{out}}+\psi_{\text{in}}=0$. 
(There is also a zero of $w$ at $\omega=0$, but it is not a pole 
because the factor $\omega$ in $w$ is canceled out by virtue of the
differentiation of Eq.~(\ref{eq:Hout}) with respect to $\log(-u)$.) 
As shown in
Appendix~\ref{sec:homogeneous} through the analysis of $\psi_{\text{out}}$
and $\psi_{\text{in}}$ near $z=z_{1}$, if the background spacetime is
dynamical, the zeros of $w(\omega)$ should change to $\omega=\pm \kappa_{1}i$ (note that the value of $\kappa_{1}$ becomes unity in
the limit to the Minkowski background). From the same analysis of the
different pair given by Eq.~(\ref{eq:barf}) near $z=z_{2}$ it is easy to
see that the outgoing mode $\bar{\psi}_{\text{out}}$ remains linearly
independent of the ingoing mode $\bar{\psi}_{\text{in}}$ at the frequencies
$\omega=\pm \kappa_{1}i$, because in general  the equality
$\kappa_{1}=\kappa_{2}$ does not hold. Hence, the requirement such
that
$0=\psi_{0}=\psi_{\text{out}}+\psi_{\text{in}}=\bar{\psi}_{\text{out}}+\bar{\psi}_{\text{in}}=0$
at $\omega=\pm \kappa_{1}i$ can become consistent only if both
$\bar{g}_{\text{out}}(z,\pm \kappa_{1}i)$ and
$\bar{g}_{\text{in}}(z,\pm \kappa_{1}i)$ vanish at any $z$ (in the
same way as $\psi_{0}$). It is also required that the coefficients $b_{1}$
and $b_{2}$ in Eq.~(\ref{eq:fout}) must be inversely proportional to the
factor $\omega^{2}+\kappa_{1}^{2}$ for keeping the functions
$g_{\text{out}}(z,\omega)$ and $g_{\text{in}}(z,\omega)$ finite even at
$\omega=\pm \kappa_{1}i$ (namely, $b_{1}\bar{g}_{\text{out}}$ and
$b_{2}\bar{g}_{\text{in}}$ become finite in the limit $\omega\rightarrow\pm\kappa_{1}i$). Then, the residue calculus of
Eq.~(\ref{eq:Hout}) for the  zeros of $w$ at $\omega=\pm\kappa_{1}i$
leads to the result that the pole contributions (denoted by
$\dot{H}^{(+)}_{\text{out}}$ and $\dot{H}^{(-)}_{\text{out}}$) to
$\dot{H}_{\text{out}}$ exist only for the source point located in the
range $|v'|<|u|<|u'|$ (namely, in the region $\mathcal{A}$ drawn in
Fig.~\ref{fg:idiagram}), and they are written by
\begin{equation}
 \dot{H}^{(\pm)}_{\text{out}}(u|u',v')=\mp
\frac{ \kappa_{1}i\lim^{\pm}\left\{b_{1}\left(\omega\right)\bar{g}_{\text{out}}\left(z_{2},\omega\right)\right\}g_{\text{out}}\left(y,\pm\kappa_{1}i\right)}
{s e^{\nu(y)-\lambda(y)}w'\left(\pm\kappa_{1}i\right)}\times\left(\frac{\left|v'\right|}{-u}\right)^{\pm \kappa_{1}},
\end{equation}
where $w'=dw/d\omega$, $\lim^{\pm}$ means the limit
$\omega\rightarrow\pm \kappa_{1}i$, and the relation 
\begin{equation}
 g_{\text{out}}\left(y,\pm\kappa_{1}i\right)|u'|^{\mp \kappa_{1}}=-g_{\text{in}}\left(y,\pm\kappa_{1}i\right)|v'|^{\mp \kappa_{1}},\label{eq:gout}
\end{equation}
is used. 

Though the physical interpretation of the modes satisfying the relation
$\psi_{\text{out}}=-\psi_{\text{in}}$ at $\omega=\pm\kappa_{1}i$ is unclear,
their contribution giving $\dot{H}^{(\pm)}_{\text{out}}$ represents a
prompt emission of outgoing waves in the region $\mathcal{A}$ including
the past Cauchy horizon. In particular, for the growing mode with
$\text{Im}(\omega)=\kappa_{1}>0$ the function $H^{(+)}_{\text{out}}$
($\propto u^{-\kappa_{1}}$) can infinitely increase as the retarded
time $u$ of the observation point approaches zero. However, the source
region in the range $u<v'<-u$ is restricted to the past Cauchy horizon
in the limit $u\rightarrow 0$, and the $\omega=\kappa_{1}i$
contribution (denoted by $\Psi_{+}$) to $\Psi(u)$ in Eq.~(\ref{eq:A}) turns
out to have the asymptotic behavior
\begin{equation}
 \Psi_{+}(u) \sim (-u)^{\kappa_{1}},
\end{equation}
because the Jacobian is given by $J(u',v') \sim (1/v')(v'/u')^{\kappa_{1}}$ near $v'=0$, 
and $g_{\text{out}}(y, \kappa_{1}i)$ is finite at $y=z_{1}$ (see
Appendix~\ref{sec:homogeneous}). This is quite similar to the high-frequency contribution $\Psi_{F1}$, which is a result of the reflection of an ingoing flux (generated on the null line $v'=u$) just at the regular center $u'=v'=u$. The $\omega=\kappa_{1}i$ contribution also may be a result of the conversion of an ingoing flux (generated in the region $\mathcal{A}$) into an outgoing flux, which occurs on the null line $u'=u$ with the finite range $u<v'<-u$. 

In the calculation of the contribution (denoted by $\Psi_{-}$) from the
decaying mode with $\text{Im}(\omega)=-\kappa_{1}<0$, 
we must remark that $g_{\text{in}}(y, -\kappa_{1}i)$ is finite at
$y=z_{1}$ (see Appendix~\ref{sec:homogeneous}), thus from
Eq.~(\ref{eq:gout}) $g_{\text{out}}(y, -\kappa_{1}i) \sim |v'/u'|^{\kappa_{1}}$ near $v'=0$.
Then, we find the asymptotic behavior such that $\Psi_{-}(u) \sim (-u)^{2\kappa_{1}}$ in the limit $u\rightarrow 0$. It is obvious that this $\omega=-\kappa_{1}i$ contribution becomes unimportant in comparison with $\Psi_{+}$.

In the above-mentioned analysis we have revealed the contributions from various source points located in the range $u\leq v'<\infty$ to the outgoing flux observed at the future null infinity. It is shown here that we can also obtain a contribution from the source point located in the range $u'\leq v'<u$ corresponding to the inner part of the region I, if there exist quasi-normal modes of the dynamical self-similar system with complex frequencies giving $w(\omega)=0$, at which the function $\psi_{\text{out}}(z,\omega)$ outgoing at $z=z_{1}$ becomes regular at the regular center $z=0$. Namely, we assume the relation $\psi_{\text{in}}=\psi_{0}-\psi_{\text{out}}=0$ at $\omega=\omega_{q} \ (q=1,2, \cdots)$, instead of the relation $\psi_{0}=0$ at $\omega=\pm\kappa_{1}i$. 

The real part of $\omega_{q}$ for such quasi-normal modes may be
nonzero, and it is proved in Appendix~\ref{sec:homogeneous} that the
imaginary part of $\omega_{q}$ should be negative. Further, from
Eqs.~(\ref{eq:fout}) and (\ref{eq:f02}),  the condition
$\psi_{\text{in}}=0$ leads to the result such that $b_{1}=b_{2}=1$. Then,
using Eq.~(\ref{eq:Hout}) and the residue calculus associated with these
poles, we obtain the $\omega=\omega_{q}$ contribution (denoted by
$\dot{H}^{(q)}_{\text{out}}$) to $\dot{H}_{\text{out}}$ as follows,
\begin{equation}
 \dot{H}^{(q)}_{\text{out}}(u|u',v')= -
\frac{\omega_{q}\bar{g}_{\text{out}}(z_{2},\omega_{q})g_{\text{out}}(y,\omega_{q})}
{s e^{\nu(y)-\lambda(y)}w'(\omega_{q})}\times\left(\frac{u}{v'}\right)^{i\omega_{q}}.\label{eq:dotHqout}
\end{equation}
Note that Eq.~(\ref{eq:dotHqout}) is valid only in the range $u'\leq v'<u$, and $\dot{H}^{(q)}_{\text{out}}=0$ at any other source points in the regions I, II and III. The quasi-normal modes represent a disturbance excited in the inner region around the regular center, which can generate outgoing waves arriving at the future null infinity. 

The key point to calculate the contribution (denoted by $\Psi_{q}$) from
the quasi-normal modes to $\Psi(u)$ is that the integral in
Eq.~(\ref{eq:A}) with respect to $v'$ is limited to the range $u'\leq
v'\leq u$, and the integrand becomes proportional to the factor
$|v'|^{-i\omega_{q}+\kappa_{1}-1}$ in the limit $v' \rightarrow
0$. Then, it is easy to see that the asymptotic behavior of $\Psi_{q}$ in
the  limit $u\rightarrow 0$ is given by
\begin{equation}
 \Psi_{q}(u) \sim (-u)^{i\omega_{q}},
\end{equation}
if the quasi-normal mode decays slowly (namely,
$|\text{Im}(\omega_{q})|<\kappa_{1}$). Interestingly, we find an
oscillatory behavior $\Psi_{q}\sim e^{i\omega_{q}\log(-u)}$ for the outgoing
flux observed at the future null infinity.
However, if it decays rapidly (namely,
$|\text{Im}(\omega_{q})|>\kappa_{1}$), we obtain again the power-law
behavior given by
\begin{equation}
 \Psi_{q}(u) \sim (-u)^{\kappa_{1}}.
\end{equation}
Then, no new feature appears for the time evolution of $\Psi(u)$, even if the quasi-normal mode is generated.

In summary, we have obtained various contributions to
$\dot{a}_{\text{out}}(u)$ representing the outgoing flux observed at the
retarded time $u$ in terms of the decomposition of the Green's function
as follows,
\begin{equation}
 \dot{a}_{\text{out}}=\Psi_{F2}+\Phi+\Psi_{F1}+\Psi_{+}+\sum_{q}\Psi_{q},\label{eq:dotaout2}
\end{equation}
where the unimportant contribution $\Psi_{-}$ is neglected. According to
the difference of the source region, they show different asymptotic
behaviors in the limit $u\rightarrow 0$. The first two terms $\Psi_{F2}$
and $\Phi$ due to outgoing waves originated on the past null cone $u'=u$
show the power-law behavior with the power index $\kappa_{2}$, while
the next two terms $\Psi_{F1}$ and $\Psi_{+}$ with the source region near the
past Cauchy horizon $v'=0$ show the power-law behavior with the power
index $\kappa_{1}$. If slowly decaying quasi-normal modes are excited
near the regular center, an oscillatory  behavior with respect to the
logarithmic time $\log|u|$ can appear in the final term $\sum_{q}\Psi_{q}$. The important point is whether or not these terms can induce a divergent energy flux in the limit $u\rightarrow 0$, at which the effect of formation of a naked singularity appears. In the next section we will define the relation of $\dot{a}_{\text{out}}$ to the outgoing energy flux measured by distant observers and discuss the values of $\kappa_{2}$, $\kappa_{1}$ and $\omega_{q}$ under specific self-similar models.

\section{The outgoing energy flux\label{sec:energy}}
In the previous sections we have used the comoving null coordinates $u$
and $v$ for specifying the observation point. The metric component
$g_{uv}$ for the double null coordinates is written by
\begin{equation}
 g_{uv} = \frac{t^{2}e^{2\nu}(V^{2}-1)}{uv}.\label{eq:guv}
\end{equation}
The comoving coordinates become crucially different from the double null
coordinates $\bar{u}$ and $\bar{v}$ used by a ``static'' observer present
at a point  sufficiently distant from the center, where we obtain
$g_{\bar{u}\bar{v}}\simeq 1$. As was shown in \cite{BarveS:NPB532:1998},
if the spacetime describing a self-similar collapse is smoothly matched
to the Schwarzschild spacetime at the star's surface, the retarded (or
advanced) time $\bar{u}$ (or $\bar{v}$) becomes equal to the retarded
(or advanced) Eddington-Finkelstein null coordinate near $\bar{u}=0$ (or
near $\bar{v}$=0). Hence, in this section we evaluate the outgoing
energy flux $F(\bar{u})$ per a unit time of $\bar{u}$, instead of $u$,
which is important as a quantity measured by the static observer near
the future null infinity.
Near the future null infinity (namely, at $z\simeq z_{2}$), Eq.~(\ref{eq:guv}) leads to $g_{uv}\sim
(|u|/v)^{\kappa_{2}-1}$, and we find the relation
\begin{equation}
 |\bar{u}| \sim |u|^{\kappa_{2}}.
\end{equation}
Hence the outgoing
energy flux $F(\bar{u})$ is given by
\begin{equation}
 F(\bar{u}) \propto \left(\frac{da_{\text{out}}}{d\bar{u}}\right)^{2}
  \sim \left(\frac{u}{\bar{u}}\right)^{2}\times\left(\frac{da_{\text{out}}}{du}\right)^{2}.\label{eq:F}
\end{equation}
It is also remarkable that near the past null infinity  corresponding to
$z\simeq z_{1}$, we obtain $|\bar{v}| \sim |v|^{\kappa_{1}}$.

It is clear from Eq.~(\ref{eq:dotaout2}) that the outgoing flux defined
by $da_{\text{out}}/du$ diverges in the limit $u\rightarrow 0$, because
in general the parameters $\kappa_{1}$ and $\kappa_{2}$ become
smaller than unity, as was mentioned in
Eq.~(\ref{eq:energycondition}). For example, the first two terms
$\Psi_{F2}$ and $\Phi$ in Eq.~(\ref{eq:dotaout2}) give the divergent term
$|u|^{\kappa_{2}-1}$ to $da_{\text{out}}/du$, which will represent a
result of prompt emission of outgoing waves (measured by the comoving
time $u$) on the past null cone $u'=u$ in the range $v'>u$, where the
coordinates $u'$ and $v'$ denote a source point for wave
generation. However, the ratio $u/\bar{u}$ of the two retarded times in
Eq.~(\ref{eq:F}) works as a red-shift factor in calculation of the
energy flux defined by $F(\bar{u})$, and the $\Psi_{F2}$ and $\Phi$
contribution (denoted by $f_{1}(\bar{u})$) to $da_{\text{out}}/d\bar{u}$ remains finite even in the limit $\bar{u}\rightarrow0$.

On the other hand the next two terms $\Psi_{F1}$ and $\Psi_{+}$ in
Eq.~(\ref{eq:dotaout2}) give the divergent term $|u|^{\kappa_{1}-1}$
to $da_{\text{out}}/du$. As was mentioned in the previous section, these
terms will represent a result of the reflection of an ingoing flux
propagating along the null line $v'=u\rightarrow 0$, where we obtain
$|\bar{v}'|\sim |v'|^{\kappa_{1}}$ for the advanced time $\bar{v}'$
of the static observer. Thus, the divergent term $|u|^{\kappa_{1}-1}$
may be interpreted as a blue-shift factor $\bar{v}'/v'$ for ingoing
waves propagating along the path $v'=u$ toward the center. Of course,
the red-shift effect also becomes important when the reflected waves
propagate to the future null infinity, and the $\Psi_{F1}$ and $\Psi_{+}$
contribution (denoted by $f_{2}$) to $da_{\text{out}}/d\bar{u}$ is given by
\begin{equation}
 f_{2}(\bar{u}) \sim
  \frac{u}{\bar{u}}\times\frac{|u|^{\kappa_{1}}}{|u|} \sim
  |\bar{u}|^{\xi_{0}}, \qquad \xi_{0}=\frac{\kappa_{1}}{\kappa_{2}}-1,\label{eq:f2}
\end{equation}
in the limit $\bar{u}\rightarrow 0$. 
If $\kappa_{2}$ is smaller than $\kappa_{1}$, 
the contribution $f_{2}$ becomes zero rather than finite as
$\bar{u}\rightarrow 0$, which contrasts with the limit of $f_{1}$.
This may mean that the red-shift of the reflected waves is
more pronounced than that of the waves without passing through the center.
 
Let us denote wave frequencies for quasi-normal modes by
$\omega_{q}=\omega_{Rq}+i\omega_{Iq}$ for real $\omega_{Rq}$ and
$\omega_{Iq}$. If $|\omega_{Iq}|$ is larger than $\kappa_{1}$ for all
$q$, the $\Psi_{q}$ contribution (denoted by $f_{3q}$) to $da_{\text{out}}/d\bar{u}$ shows the
same power-law behavior as $f_{2}$. We note that the amplification due
to the blue-shift effect on the null line $v'=u$ can still work even for
rapidly decaying quasi-normal modes. If $|\omega_{Iq}|$ becomes smaller
than $\kappa_{1}$ for some $q$, the contribution $f_{3q}$
in the limit $\bar{u}\rightarrow 0$ can be written by the oscillatory
form
\begin{equation}
 f_{3q}(\bar{u}) \sim
  |\bar{u}|^{\xi_{q}}\cos\left(\frac{\omega_{Rq}}{\kappa_{2}}\log|\bar{u}|\right),
  \qquad \xi_{q}=\frac{|\omega_{Iq}|}{\kappa_{2}}-1.\label{eq:f3q}
\end{equation}
Owing to prompt generation of waves induced by slowly decaying quasi-normal modes in the range $u'<v'<u$, the blue-shift effect on the null line $v'=u$ becomes insignificant.

In conclusion, we find the asymptotic behavior (in the limit
$\bar{u}\rightarrow 0$) of the outgoing energy flux $F(\bar{u})$
measured by static observers present near the future null infinity as
follows,
\begin{equation}
 F(\bar{u})\propto\left(f_{1}(\bar{u})+f_{2}(\bar{u})+\sum_{q}f_{3q}(\bar{u})\right)^{2}
  \simeq\left\{c_{1}+c_{2}|\bar{u}|^{\xi_{0}}+\sum_{q}c_{3}|\bar{u}|^{\xi_{q}}\cos\left(\frac{\omega_{Rq}}{\kappa_{2}}\log|\bar{u}|\right)\right\}^{2},\label{eq:F2}
\end{equation}
where $c_{1}$, $c_{2}$ and $c_{3}$ are constants. The third term in
Eq.~(\ref{eq:F2}) appears only if there exist quasi-normal modes
satisfying the condition $|\omega_{Iq}|<\kappa_{1}$. By virtue of the
red-shift effect no divergence of the energy flux $F(\bar{u})$ occurs
even at the moment when the effect of naked singularity formation
arrives at the future null infinity, unless $\kappa_{1}$ or
$|\omega_{Iq}|$ becomes smaller than $\kappa_{2}$. For the self-similar
collapse such that $\kappa_{1}$ becomes smaller than $|\omega_{Iq}|$
and $\kappa_{2}$, the power-law divergence
\begin{equation}
 F(\bar{u}) \sim |\bar{u}|^{2\xi_{0}},
\end{equation}
will be observed as a dominant behavior with the power index limited to
the range $-2<2\xi_{0}<0$. If slowly decaying quasi-normal modes are
excited (namely, if $|\omega_{Iq}|$ is smaller than $\kappa_{1}$ and
$\kappa_{2}$), the oscillatory divergence
\begin{equation}
 F(\bar{u}) \sim
  |\bar{u}|^{2\xi_{q}}\cos^{2}\left(\frac{\omega_{Rq}}{\kappa_{2}}\log|\bar{u}|\right),
  \qquad -2<2\xi_{q}<0,
\end{equation} 
can become a more interesting precursor of naked singularity formation. Of course, the generation of a divergent energy flux means that the naked singularity formation is an unstable process against a backreaction of perturbations. Nevertheless the above-mentioned precursor phenomenon may significantly appear, before the backreaction effect becomes important. Even if $F(\bar{u})$ remains finite at $\bar{u}=0$, the anomalous time evolution of $F(\bar{u})$ due to the damping term $|\bar{u}|^{\xi_{0}}$ or $|\bar{u}|^{\xi_{q}}\cos(\omega_{Rq}\log|\bar{u}|/\kappa_{2})$ will appear in principle as an observable effect different from usual black hole formation. 

Here we briefly comment on the outgoing energy flux in black hole
formation in terms of Eq.~(\ref{eq:F2}). This corresponds to the case
$V'(z_{2})\rightarrow 0$ (namely, $z_{m}=z_{2}=z_{3}$ in Fig.~\ref{fg:V}), for which the Cauchy horizon coincides with the
event horizon. Then, the power indices $\xi_{0}$ and $\xi_{q}$ become
infinitely large, and the power law decay of the second and third terms
in Eq.~(\ref{eq:F2}) will change to an exponential decay. The first term
$c_{1}$ in Eq.~(\ref{eq:F2}) also vanishes in the limit
$V'(z_{2})\rightarrow 0$, because the Jacobian $J(u,v')$ involved in the
integrals (\ref{eq:AF}) and (\ref{eq:B}) (giving $\Psi_{F2}$ and $\Phi$) is
found to be proportional to the parameter $V'(z_{2})$ at the leading order
in the limit $u\rightarrow 0$, by considering the constant factor
omitted in Eq.~(\ref{eq:J2}). The formula given by Eq.~(\ref{eq:F2}) can
describe the infinite red-shift effect at the event horizon which makes
the outgoing energy flux  $F(\bar{u})$ completely vanish at $\bar{u}=0$. 

\begin{figure}
 \includegraphics[height=7cm]{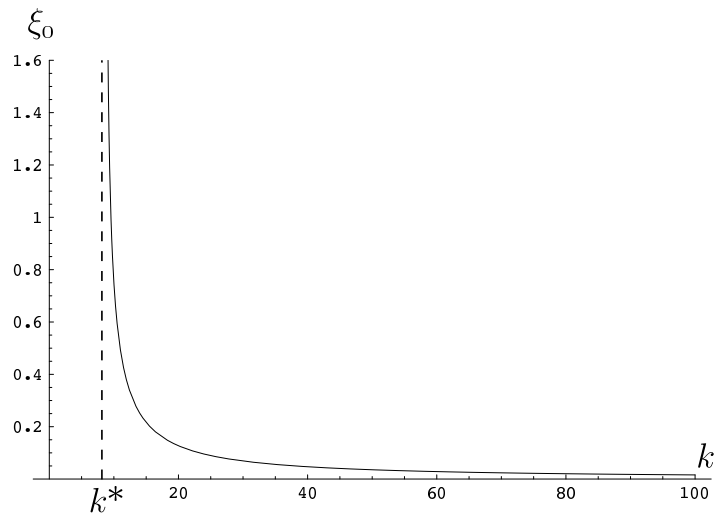}
 \caption{Numerical evaluation of the power law index $\xi_{0}$ for dust collapse parameterized by $k$. The value of $\xi_{0}$ diverges at $k=k^{*}$, where the future Cauchy horizon coincides with the event horizon.}
 \label{fg:dust}
\end{figure}

The key parameters for determining the asymptotic evolution of
$F(\bar{u})$ are the ratios $\kappa_{1}/\kappa_{2}$ and
$|\omega_{Iq}|/\kappa_{2}$. To discuss the values of the parameters, we
must specify the self-similar model. If the collapsing matter is a
perfect fluid, the self-similarity (\ref{eq:self-similarity}) is
satisfied only for the equation of state $P=\alpha \rho$, and we find
the inequality $\kappa_{1}>\kappa_{2}$ for $0\leq\alpha\leq 1$. The
general proof of this inequality except the case $\alpha=0$ is given in
Appendix~\ref{sec:perfect}. For the collapse of dust (namely, for
$\alpha=0$) we can obtain explicitly the metric (\ref{eq:metric1}) as
follows,
\begin{equation}
 \nu(z)=0, \qquad  S(z)=\left\{\frac{3}{2}\left(k-\frac{1}{z}\right)\right\}^{2/3},
\end{equation}
from which the function $V(z)$ is derived as
\begin{equation}
 V(z)=ze^{\lambda}=\frac{1}{3}\left\{\frac{9z}{4(k
		  z-1)}\right\}^{1/3}\left(3k
		  z-1\right),
\end{equation}
where $k$ is a positive constant. The global structure described in
Fig.~\ref{fg:diagram} (corresponding to the behavior of $V$ shown in
Fig.~\ref{fg:V}) is possible only in the range
$k>k^{*}=(26+15\sqrt{3})/6\simeq 8.7$, and from Fig.~\ref{fg:dust} it is easy
to see that the power law index $\xi_{0}$ becomes always positive. 
Naked singularity formation due to self-similar collapse of a perfect fluid will not induce a  divergent outgoing energy flux $F(\bar{u})$, unless there exist slowly decaying quasi-normal modes with $|\omega_{Iq}|/\kappa_{2}<1$. It will be interesting to check whether or not the inequality $\kappa_{1}>\kappa_{2}$ holds for any other collapsing matter (for example, in the case of a scalar-field collapse).

To analyze the quasi-normal modes in general, it will be useful to
rewrite the homogeneous equation for the outgoing mode
$\psi_{\text{out}}(z,\omega)$ into an one-dimensional Schr\"{o}dinger-type
differential equation with a potential $U(z)$, which is given by the
metric components $\nu$, $S$ and $\lambda$ and decreases in
proportion to $1-V^{2}$ in the limit $z\rightarrow z_{1}$ (see
Appendix~\ref{sec:homogeneous} for the equation). The existence of the
quasi-normal solution $\psi_{\text{out}}$ satisfying the regularity at
$z=0$ is shown in Appendix~\ref{sec:quasi}, by assuming some suitable
form of $U$. We find the case such that the absolute value
$|\omega_{Iq}|$ of the imaginary part of the complex frequency
$\omega_{q}$ with nonzero $\omega_{Rq}$ can be smaller than $\kappa_{1}$. However, it remains unclear whether or not such a case is possible for the metric satisfying the Einstein equations. This problem would be more extensively investigated in future works.

We have developed the Green's function technique to study time evolution of electromagnetic perturbations in the self-similar background, which will be applicable to any other test fields (for example, scalar and gravitational fields). We have also clarified that the various contributions to the Green's function (for example, high frequency part and quasi-normal modes) given by Eq.~(\ref{eq:dotaout2}) are closely related to the difference of the source region where the corresponding perturbations are generated. 

Our main result  is the finding of the key parameters
$\kappa_{1}/\kappa_{2}$ and $|\omega_{Iq}|/\kappa_{2}$ to determine the
occurrence of divergent energy flux associated with naked singularity
formation. The value of the complex quasi-normal frequency $\omega_{q}$
should depend not only on the background geometry but on the test field
(and $l$ parameterizing the field angular momentum). On the other hand,
the values of $\kappa_{1}$ and $\kappa_{2}$ are purely geometrical
quantities given by the derivative of the metric on the past and future
Cauchy horizons, respectively. The blue-shift and red-shift effects
defined by $\bar{v}/v$ and $u/\bar{u}$ for ingoing and outgoing waves
propagating near the Cauchy horizons are parameterized by these
geometrical quantities. The competition of such two effects in
generation of outgoing energy flux would be an important feature of
naked singularity.

It is interesting to compare our result with the result
giving by Chandrasekhar and Hartle\cite{SChandrasekhar:PRSLA384:1982},
who have considered time
evolution of electromagnetic and gravitational perturbations inside the 
Reissner-Nordstr\"{o}m black hole. The black hole contains a timelike singularity
visible by a falling observer in the black hole.
Estimating the energy flux received
by an observer near the (future) Cauchy horizon for the timelike
singularity, 
they have found that its amplitude has an exponential behavior  
with the rate $\kappa_{-}-\kappa_{+}$, where $\kappa_{+}$ and
$\kappa_{-}$ are the surface gravity factors evaluated on the event
horizon and the Cauchy horizon, respectively (note that the event
horizon corresponds to the past Cauchy horizon).
Because the rate $\kappa_{-}-\kappa_{+}$ is always positive, 
the energy flux diverges when the observer arrives at the Cauchy horizon.
The power index $\xi_{0}$, which is written
by the difference of the constant $\kappa$ evaluated on the
future and past Cauchy horizon by Eq.~(\ref{eq:chi}), resembles the
rate, except that it is
not always positive.
This commonality would provide an important insight into 
an universal feature of perturbation response
in various geometries with a naked singularity. 

Further we would like to emphasize that the power index $\gamma$ of the divergent energy flux $F\sim |\bar{u}|^{\gamma}$ with (or without) an oscillation (corresponding to $\gamma=2\xi_{q}$ (or $\gamma=2\xi_{0}$)) is allowed in the range $-2<\gamma <0$. This sharply contrasts with the semiclassical result giving the unique power index $\gamma=-2$ for the radiation of quantized test fields without depending on the parameters of the self-similar background\cite{HiscockWA:PRD26:1982, BarveS:NPB532:1998, BarveS:PRD58:1998, VazC:PLB442:1998, SinghTP:PLB481:2000.2, MiyamotoU:PRD69:2004}. Such a difference of the power-law divergence may become an important clue for understanding profoundly quantum properties of perturbations around naked singularity.

\appendix

\section{Application to Minkowski background \label{sec:Minkowski}}
Let us apply Eq.~(\ref{eq:a4}) to Minkowski background $\nu=\lambda=0$
and $S=1$ for checking the validity of the formulas given in
Sec.~\ref{sec:Green}. Here we consider the dipole field with
$l=1$. Then, the
homogeneous solutions for Eq.~(\ref{eq:ODE}) are easily given by
\begin{equation}
 g_{\text{out}}(z,\omega)=\omega-\frac{i}{z}, \qquad g_{\text{in}}(z,\omega)=\omega+\frac{i}{z}.
\end{equation}
Of course we obtain $b_{1}=b_{2}=1$ for the coefficients in
Eq.~(\ref{eq:fout}), and the Wronskian factor is given by
\begin{equation}
 w(\omega) = 2i\omega(\omega+i)(\omega-i).
\end{equation}
Then, from Eqs.~(\ref{eq:tildeG}), (\ref{eq:H1}) and (\ref{eq:H2}) we
find
\begin{equation}
 H(u,v|u',v')=\frac{2(uv+u'v')-(v'+u')(v+u)}{(v-u)(v'-u')(v'+u')},
\end{equation}
for $u<v'$, and $H(u,v|u',v')=0$ for $v'<u$. To calculate explicitly
Eq.~(\ref{eq:a4}), we assume the shell-like current distribution
$j_{1}(t,z)=j_{0}\delta(r-r_{0})$ for a constant $j_{0}$. Then, we find
$a(r)=-4\pi j_{0}r^{2}_{0}/3r$ for $r>r_{0}$, and $a(r)= -4\pi
j_{0}r^{2}/3r_{0}$ for $r<r_{0}$. This corresponds to the vector
potential for a static dipole magnetic field.

\section{Analysis of homogeneous solutions \label{sec:homogeneous}}
Here we analyze the homogeneous solutions $\psi(z,\omega)$ for
Eq.~(\ref{eq:ODE}), by rewriting it into the one-dimensional
Schr\"{o}dinger-type differential equation
\begin{equation}
 \frac{d^{2}\tilde{\psi}}{dx^{2}}+\left(\omega^{2}-U(x)\right)\tilde{\psi}=0,\label{eq:ODE2}
\end{equation}
where the new variable $x$ and the ``potential'' $U$ are given by
\begin{equation}
 x=\frac{1}{2}\left(h_{\text{out}}(z)-h_{\text{in}}(z)\right),
\end{equation}
and
\begin{equation}
 U(x) = \frac{l(l+1)e^{2\nu}(1-V^{2})}{S^{2}z^{2}}.\label{eq:U}
\end{equation}
Note that $U>0$ in the region I shown in Fig.~\ref{fg:diagram}. The
function $\tilde{\psi}$ is defined by
\begin{equation}
 \tilde{\psi}(x,\omega)\equiv \psi(z,\omega)e^{-i\omega(h_{\text{out}}(z)+h_{\text{in}}(z))/2}.\label{eq:tildef}
\end{equation}
To obtain the homogeneous solutions giving $\psi_{0}$ and $\psi_{\text{out}}$,
we must consider the behavior at the points $z=0$ and $z=z_{1}$
corresponding to $x=0$ and $x=\infty$, respectively. The local flatness
at the regular center and the suitable gauge choice allow us, without
loss of generality, to assume that $\nu\simeq 0$ and $e^{\lambda}\simeq mS \simeq C_{\lambda}(-z)^{m-1}$ near $z=0$\footnote{Then, via the
coordinate transformation $\hat{r}=r^{1+m}$, we can find that near
$z=0$,\[ds^{2} \simeq
      -dt^{2}+\left\{\frac{C_{\lambda}(-t)^{1-m}}{m}\right\}^{2}(d\hat{r}^{2}+\hat{r}^{2}d\Omega^{2}).\]},
 where $C_{\lambda}$ and $m$ are positive constants. Then, we find
the approximate form
\begin{equation}
 U(x) \simeq \frac{l(l+1)}{x^{2}},\label{eq:asymU2}
\end{equation}
near $x=0$. On the other hand it is easy to see that
\begin{equation}
 U(x) \simeq U_{\infty}e^{-2\kappa_{1}x},\label{eq:asymU}
\end{equation}
in the limit $x\rightarrow\infty$, where $U_{\infty}$ is a positive constant. It is remarkable that we have $U=l(l+1)/\sinh^{2}x$ and $z=-\tanh x$ for the Minkowski metric.

Denoting the solutions for Eq.~(\ref{eq:ODE2}) by
$\tilde{\psi}_{\text{out}}$ and $\tilde{\psi}_{\text{in}}$ corresponding to
$\psi_{\text{out}}$ and $\psi_{\text{in}}$, respectively, we obtain
\begin{equation}
 \tilde{\psi}_{\text{out}}(x,\omega)= g_{\text{out}}(z,\omega)e^{i\omega x}, \qquad \tilde{\psi}_{\text{in}}(x,\omega)= g_{\text{in}}(z,\omega)e^{-i\omega x}.
\end{equation}
The approximate forms of these functions for large $x$ are given by
\begin{equation}
 g_{\text{out}}(x,\omega) \simeq g^{\infty}_{\text{out}}(\omega)\left\{1+\frac{U_{\infty}}{4\kappa_{1}(\kappa_{1}-i\omega)}e^{-2\kappa_{1}x}\right\},
\end{equation}
and
\begin{equation}
 g_{\text{in}}(x,\omega) \simeq g^{\infty}_{\text{in}}(\omega)\left\{1+\frac{U_{\infty}}{4\kappa_{1}(\kappa_{1}+i\omega)}e^{-2\kappa_{1}x}\right\},
\end{equation}
for some functions $g^{\infty}_{\text{out}}$ and
$g^{\infty}_{\text{in}}$ dependent on $\omega$.
To keep these functions well-defined even at $\omega=\pm\kappa_{1}i$, 
the functions $g^{\infty}_{\text{out}}$ and $g^{\infty}_{\text{in}}$ are
required to contain the factor $\kappa_{1}-i\omega$ and
$\kappa_{1}+i\omega$, respectively.
Then, we obtain $w(\omega)=2i\omega g^{\infty}_{\text{out}}g^{\infty}_{\text{in}}=0$ at $\omega=\pm\kappa_{1}i$, for which it is easy to see that the two modes $\tilde{\psi}_{\text{out}}$ and $\tilde{\psi}_{\text{in}}$ become linearly dependent at the leading order in the limit $x\rightarrow\infty$. 

Next, using Eq.~(\ref{eq:ODE2}), we prove that the imaginary  part of
the frequencies $\omega_{q}$ of quasi-normal modes denoted by
$\tilde{\psi}_{q}(x)$ should be negative. We require the boundary conditions
for the quasi-normal modes such that $\tilde{\psi}_{q}\sim x^{l+1}$ near
$x=0$, and $\tilde{\psi}_{q}\sim e^{i\omega_{q}x}$ in the limit
$x\rightarrow\infty$. Then, integrating Eq.~(\ref{eq:ODE2}) multiplied
by the complex conjugate $\tilde{\psi}^{*}_{q}$ in the range $0\leq x\leq
x_{0}$, where $x_{0}$ is a positive constant satisfying
$\tilde{\psi}_{q}(x_{0},\omega_{q})\neq 0$, for sufficiently large $x_{0}$ we obtain
\begin{equation}
 X\omega_{q}^{2}+i\omega_{q}Y-Z=0,
\end{equation}
where the coefficients given by
\begin{equation}
 X=\int_{0}^{x_{0}}\left|\tilde{\psi}_{q}\right|^{2}dx, \qquad
  Y=\left|\tilde{\psi}_{q}(x_{0})\right|^{2}, \qquad Z=\int_{0}^{x_{0}}\left(\left|\frac{d\tilde{\psi}_{q}}{dx}\right|^{2}+U\left|\tilde{\psi}_{q}\right|^{2}\right)dx,
\end{equation}
are positive definite. It is easy to see that
$\text{Im}(\omega_{q})=-Y/2X<0$ for $4XZ>Y^{2}$, and
$\text{Im}(\omega_{q})=-(Y\pm\sqrt{Y^{2}-4XZ})/2X<0$ for $4XZ<Y^{2}$.

\section{Perfect fluid collapse with pressure \label{sec:perfect}}
\begin{figure}
 \includegraphics[height=7cm]{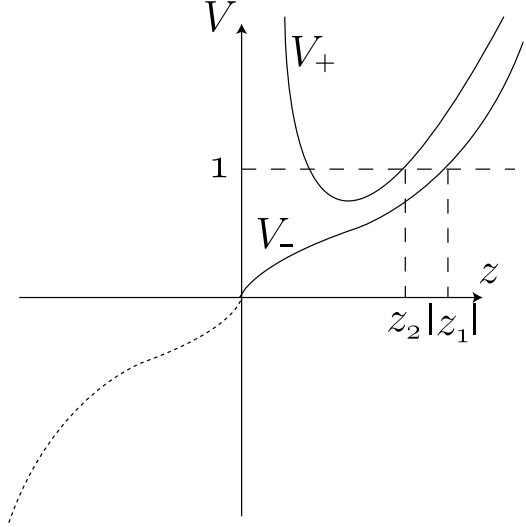}
 \caption{Schematic description (sold lines) of the functions $V_{+}(z)$ and $V_{-}(z)$. The dotted line denotes the function $V(z)$ for $z<0$.
 The line $V=V_{-}(z)$ is the point symmetric line to the dotted line
 with respect to the origin.}
 \label{fg:VV}
\end{figure}

In this appendix, we prove that 
the value of $\kappa_{2}$ is always smaller than that of
$\kappa_{1}$ in the spacetime describing the collapse of a perfect
fluid with the equation of state $P=\alpha\rho$ for $0<\alpha\leq1$.
For convenience, we express the function $V(z)$ for $z>0$ and the point
symmetric function to the function $V(z)$ for $z<0$ concerning the
origin $(z,V)=(0,0)$ by the notation $V_{+}(z)$ and $V_{-}(z)$,
respectively (see Fig.~\ref{fg:VV}).
As stated in \cite{BJCarr:PRD62:2000}, the function $V_{-}(z)$ is also
given by the
metric satisfying the Einstein equations.
In addition, we define $\dot{}\equiv zd/dz$ only in this appendix.
Hence, what we show is that the value of $\dot{V}_{+}(z_{2})$ is
always smaller than that of $\dot{V}_{-}(|z_{1}|)$.

Now we start the proof with a similar notation used by Carr and Coley in \cite{BJCarr:PRD62:2000}. 
They introduced the functions $A(z)$ and $B(z)$ defined by
$(4\pi\rho r^{2})^{-\alpha/(1+\alpha)}=A_{0}e^{A(z)}$ and $S(z)=B_{0}e^{B(z)}$ for
positive constants $A_{0}$ and $B_{0}$.
From the Einstein equations, the functions $\nu(z)$ and $\lambda(z)$
can be expressed in terms of the functions $A(z)$ and $B(z)$.
Hence, 
the function $V(z)$ for $z>0$ can be expressed as
\begin{equation}
 V(z)=V_{0}B_{0}^{-2}z^{(1-\alpha)/(1+\alpha)}e^{-2B+A(1-\alpha)/\alpha}, \label{eq:VP}
\end{equation}
for a positive constant $V_{0}$ (see Eq.~(2.19) of \cite{BJCarr:PRD62:2000}).
In addition, we use one of the Einstein equations
\begin{equation}
 V^{2}\left(\dot{B}-\frac{\dot{A}}{2\alpha}\right)=-\frac{\dot{A}}{2}+\left(\frac{\alpha}{1+\alpha}\right)\left\{e^{-4B+A(1-\alpha)/\alpha}-1\right\}, \label{eq:EE}
\end{equation}
which is shown in Eq.~(4.23) of \cite{BJCarr:PRD62:2000}.
When $V=1$, Eqs.~(\ref{eq:VP}) and (\ref{eq:EE}) give
\begin{equation}
 \dot{V}=1-\frac{2\alpha}{1+\alpha}e^{-4B+A(1-\alpha)/\alpha}.
\end{equation}
This means that the values of $\dot{V}_{-}(z)$ and $\dot{V}_{+}(z)$ at $z=|z_{1}|$
and $z=z_{2}$, respectively, become
\begin{equation}
 \dot{V}_{-}\left(|z_{1}|\right)=1-\frac{2\alpha}{V_{0}(1+\alpha)}|z_{1}|^{-(1-\alpha)/(1+\alpha)}S^{-2}(z_{1}),\label{eq:dotV+}
\end{equation}
and 
\begin{equation}
 \dot{V}_{+}(z_{2})=1-\frac{2\alpha}{V_{0}(1+\alpha)}z_{2}^{-(1-\alpha)/(1+\alpha)}S^{-2}(z_{2}).\label{eq:dotV-}
\end{equation}
As stated in \cite{BJCarr:PRD62:2000}, the solution describes the monotonical
collapse.
This means that the value of $S(z_{1})$ is larger than that of
$S(z_{2})$.
Hence, when $\alpha=1$, we find the inequality
$\dot{V}_{+}(z_{2})<\dot{V}_{-}\left(|z_{1}|\right)$ from
Eqs.~(\ref{eq:dotV+}) and (\ref{eq:dotV-}).
On the other hand, for $0<\alpha<1$, Eqs.~(\ref{eq:dotV+}) and (\ref{eq:dotV-}) 
indicate that if the value $|z_{1}|$ is larger than the value $z_{2}$, the value of $\dot{V}_{+}(z_{2})$ is
always smaller than that of $\dot{V}_{-}(|z_{1}|)$.
The condition $|z_{1}|>z_{2}$ is satisfied if the inequality
$V_{+}(z)>V_{-}(z)$ holds for the whole region $z\geq |z_{1}|$, as
depicted in Fig.~\ref{fg:VV}.
Therefore, the remaining problem is to show the inequality
$V_{+}(z)>V_{-}(z)$ in the whole region $z\geq |z_{1}|$ for $0<\alpha<1$.

We firstly show that $V_{+}(z)>V_{-}(z)$ holds at the sufficiently large $z$.
By requiring the left hand side of Eq.~(\ref{eq:EE}) to be finite as
$V\rightarrow \infty$, we find that at the sufficiently large $z$,
Eq.~(\ref{eq:EE}) becomes
\begin{equation}
 \dot{B}=\frac{\dot{A}}{2\alpha},
\end{equation} 
which is also shown in Eq.~(4.25) of \cite{BJCarr:PRD62:2000}.
From this equation and Eq.~(\ref{eq:VP}), we can obtain
\begin{equation}
 \frac{\dot{V}}{V}=\frac{1-\alpha}{1+\alpha}-2\alpha\dot{B}.
\end{equation}
Hence, we find
\begin{equation}
 \frac{V'_{-}}{V_{-}}-\frac{V'_{+}}{V_{+}}=-2\alpha(B'_{-}-B'_{+}), 
\end{equation}
where the functions $B_{-}(z)$ and $B_{+}(z)$ denote the values of
$B(z)$ along the lines of $V(z)=V_{-}(z)$ and $V(z)=V_{+}(z)$,
respectively.
The monotonical collapse also means that $B'_{-}<0$ and $B'_{+}>0$.
In addition, the equality $V_{+}(z)=V_{-}(z)$ holds in the limit
$z\rightarrow \infty$.
These facts means that the inequality $V'_{-}(z)>V'_{+}(z)$ holds.
Hence, the inequality $V_{+}(z)>V_{-}(z)$ holds at the
sufficiently large $z$.

Next, we show that the lines $V=V_{+}(z)$ and $V=V_{-}(z)$ do not cross
each other in the region $z>|z_{1}|$ to complete the proof.
We eliminate the functions $A(z)$ and $\dot{A}(z)$ from
Eq.~(\ref{eq:EE}), using the function $V(z)$ and $\dot{V}(z)$, and 
obtain 
\begin{equation}
 \frac{\alpha}{1-\alpha}\left(1-V^{2}\right)\dot{B}=\frac{\alpha}{1+\alpha}\left(V_{0}^{-1}B_{0}^{2}z^{(\alpha-1)/(\alpha+1)}Ve^{-2B}-1\right)-\frac{1}{2}\left(\alpha-V^{2}\right)\left(\frac{\dot{V}/V}{1-\alpha}-\frac{1}{1+\alpha}\right).\label{eq:EEb}
\end{equation}
We consider the difference between Eq.~(\ref{eq:EEb}) expressed in terms
of $V_{-}(z)$ and $B_{-}(z)$ and Eq.~(\ref{eq:EEb}) expressed in terms
of $V_{+}(z)$ and $B_{+}(z)$.
Then, for the region $V_{+}\geq 1$ and $V_{-}\geq 1$, we can find the
inequality
\begin{eqnarray}
 &&\left(V_{-}-\frac{\alpha}{V_{-}}\right)\dot{V}_{-}-\left(V_{+}-\frac{\alpha}{V_{+}}\right)\dot{V}_{+}\nonumber\\
&&\hspace{2cm}>\frac{1-\alpha}{1+\alpha}\left\{V_{-}^{2}-V_{+}^{2}
+2\alpha
V_{0}^{-1}B_{0}^{2}z^{(\alpha-1)/(\alpha+1)}\left(V_{+}e^{-2B_{2}}-V_{-}e^{-2B_{1}}\right)\right\}, \label{eq:difference}
\end{eqnarray}
because of the conditions $\dot{B}_{-}<0$ and $\dot{B}_{+}>0$.
Now we assume that the lines $V=V_{+}(z)$ and $V=V_{-}(z)$ cross each
other (i.e., $V_{+}(z)=V_{-}(z)=V_{c} \geq 1$) at some points $z$.
At this point, the inequality (\ref{eq:difference}) becomes
\begin{equation}
 \left(V_{c}-\frac{\alpha}{V_{c}}\right)\left(\dot{V}_{-}-\dot{V}_{+}\right)>\frac{2
  V_{0}^{-1}B_{0}^{2}V_{c}\alpha(1-\alpha)}{1+\alpha}z^{(\alpha-1)/(\alpha+1)}\left(e^{-2B_{+}}-e^{-2B_{-}}\right). 
\end{equation}
Because of the monotonical collapse condition $B_{-}(z)>B_{+}(z)$, we obtain the inequality 
$\dot{V}_{-}(z)>\dot{V}_{+}(z)$.
However, since the inequality $V_{+}(z)>V_{-}(z)$ holds at the sufficiently
large $z$, the derivatives should satisfy
$\dot{V}_{+}(z)\geq\dot{V}_{-}(z)$ 
at the point where $V_{+}(z)=V_{-}(z)$. 
This is a contradiction, which is solved by denying the existence of the
point where $V_{+}(z)=V_{-}(z)$.
Hence, the inequality $V_{+}(z)>V_{-}(z)$ holds in the whole region
$V_{-}(z)\geq 1$.
This conclusion ends the proof that the inequality
$\dot{V}_{+}(z_{2})<\dot{V}_{-}(|z_{1}|)$ always holds for
$0<\alpha\leq1$.

\section{Approximate evaluation of frequencies of quasi-normal modes
 \label{sec:quasi}}
The oscillatory behavior shown by Eq.~(\ref{eq:f3q}) can appear if there
exist quasi-normal modes with the frequencies $\omega_{q}$ satisfying
the conditions $\text{Re}(\omega_{q})\neq 0$ and
$|\text{Im}(\omega_{q})|<\kappa_{1}$, though it depends on the value
of $\kappa_{2}$ whether or not the divergence of the energy flux
occurs. In general it will be a difficult task to find quasi-normal
modes as homogeneous solutions for Eq.~(\ref{eq:ODE}) constructed by
self-similar metrics satisfying the Einstein equations. Even the
existence of such modes will be unclear. Hence, in this appendix we
consider a simple form of the potential $U(x)$ defined by
Eq.~(\ref{eq:U}). Our purpose is to illustrate that only a slight modification of $U$ from the potential given by the Minkowski metric is sufficient for obtaining quasi-normal modes. 

Taking account of the asymptotic behaviors given by
Eqs.~(\ref{eq:asymU2}) and (\ref{eq:asymU}), we can give a plausible approximation
of $U$ by
\begin{equation}
 U(x)=\left\{
\begin{array}{ll}
\displaystyle{\frac{l(l+1)}{x^{2}}} &
 \mbox{for $x\leq x_{c}$},\\
&\\
\displaystyle{\frac{\zeta+\left(U_{\infty}-\zeta\right)\left(1-e^{-2\kappa_{1}x}\right)}{4\sinh^{2}\left(\kappa_{1}x\right)}}&
 \mbox{for $x\geq x_{c}$}, 
\end{array}
 \right.\label{eq:apU}
\end{equation}
where $\zeta$ and $x_{c}$ are constants. Because the potential should
be continuous at $x=x_{c}$, the parameter $\zeta$ is given by
\begin{equation}
 \zeta = e^{2\kappa_{1}x_{c}}\left\{\frac{4l(l+1)\sinh^{2}\left(\kappa_{1}x_{c}\right)}{x_{c}^{2}}-U_{\infty}\left(1-e^{-2\kappa_{1}x_{c}}\right)\right\}.
\end{equation}
Therefore the potential $U$ given by Eq.~(\ref{eq:apU}) is parameterized by $l$, $x_{c}$, $U_{\infty}$ and $\kappa_{1}$.

\begin{figure}
\includegraphics[height=5cm]{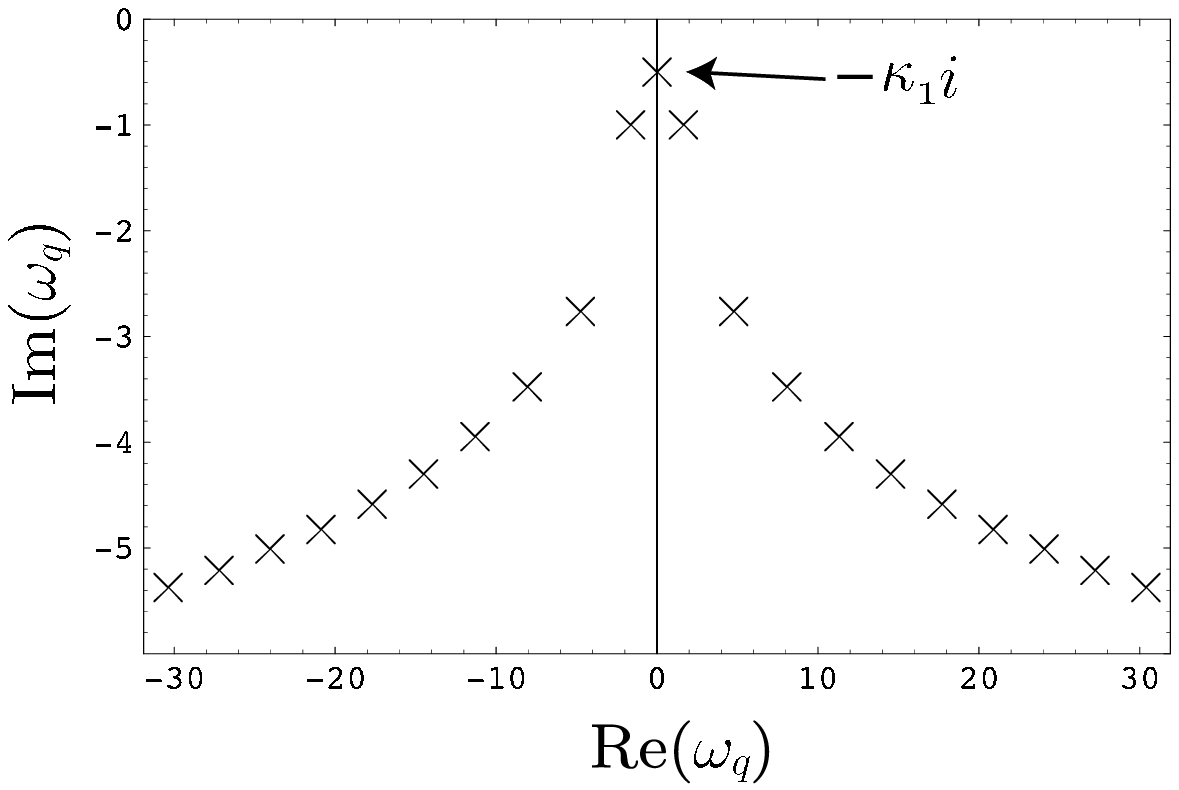}
\includegraphics[height=5cm]{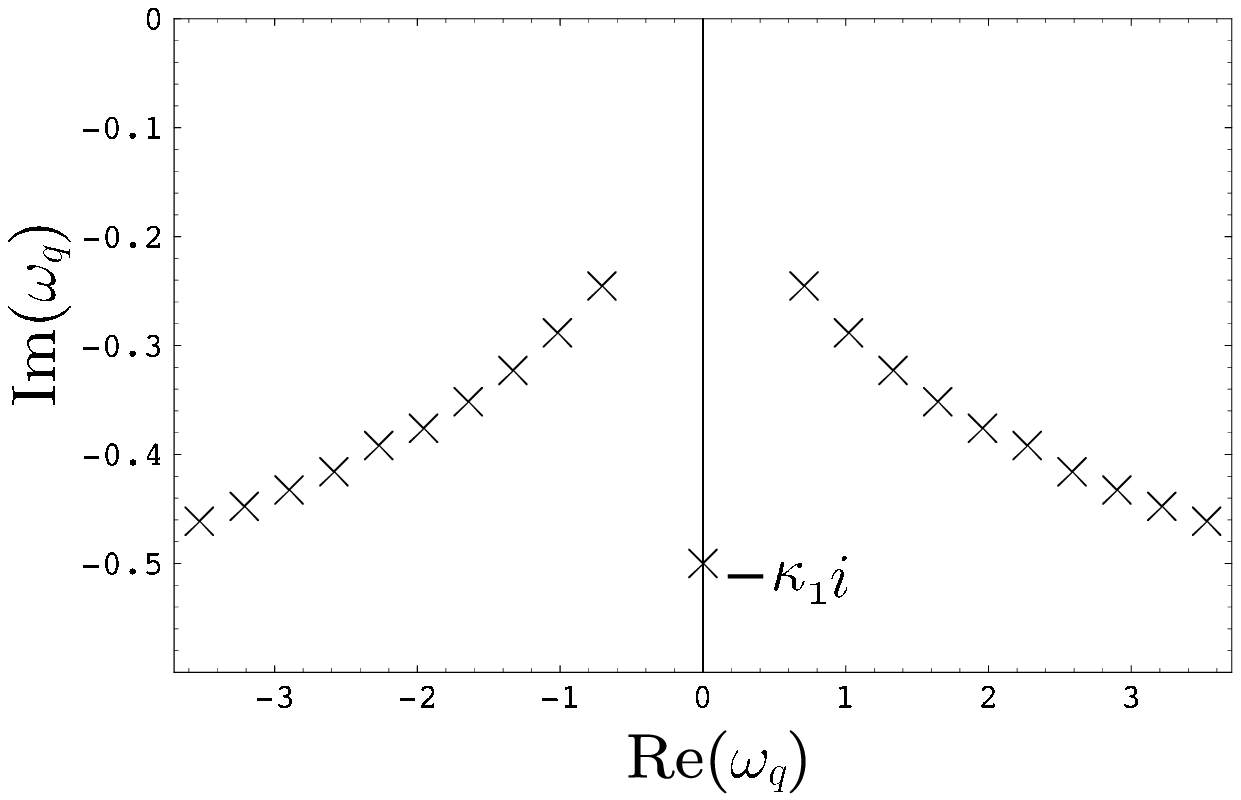}
\caption{The complex frequencies $\omega_{q}$ of quasi-normal modes for the potential $U(x)$ given by Eq. (D1). The left and  right panels correspond to the cases $x_{c}=1$ and $x_{c}=10$, respectively, while the other parameters are fixed to be $l=1$, $U_{\infty}=8$ and $\kappa_{1}=0.5$.}
\label{fg:pole}
\end{figure}

From the regularity of quasi-normal modes denoted by $\tilde{\psi}_{q}$ at
$x=0$, we obtain for $x\leq x_{c}$
\begin{equation}
 \tilde{\psi}_{q}(x)=\sqrt{x}J_{l+(1/2)}(\omega_{q}x),\label{eq:fq1}
\end{equation}
where $J_{l+(1/2)}$ denotes the Bessel function. On the other hand, from
the asymptotic behavior $\tilde{\psi}_{q}\sim e^{i\omega_{q} x}$, we obtain
for $x\geq x_{c}$
\begin{equation}
 \tilde{\psi}_{q}(x)=g^{\infty}_{\text{out}}(\omega_{q})\left(1-e^{-2\kappa_{1}x}\right)^{\mu}F\left(\beta_{1},
						      \beta_{2},1-\frac{i\omega}{\kappa_{1}};
						      e^{-2\kappa_{1}x}\right)e^{i\omega_{q}x},\label{eq:fq2}
\end{equation} 
where $\mu=\{1+\sqrt{1+(\zeta/\kappa_{1})}\}/2$, and $F$ is the
hypergeometric function with $\beta_{1}$ and $\beta_{2}$ given by
\begin{equation}
 \beta_{1}\beta_{2}=\frac{U_{\infty}}{4\kappa_{1}^{2}}-\mu\left(\frac{i\omega}{\kappa_{1}-1}\right),
\end{equation}
and
\begin{equation}
 \beta_{1}+\beta_{2}=2\mu-\frac{i\omega}{\kappa_{1}}.
\end{equation}
By requiring that the functions given by Eqs.~(\ref{eq:fq1}) and
(\ref{eq:fq2}) are smoothly matched at $x=x_{c}$, we can  find the
frequencies $\omega_{q}$. The numerical results are illustrated in
Fig.~\ref{fg:pole}. 
We can confirm that there are a number of quasi-normal modes  with
$\text{Re}(\omega_{q})\neq 0$. Further the dependence of the minimum
value (denoted by $|\text{Im}(\omega_{0})|$) of $|\text{Im}(\omega_{q})|$ on the
parameters can be summarized as follows. (1) The value of
$|\text{Im}(\omega_{0})|$ decreases as $x_{c}$ increases, and it becomes
equal to $\kappa_{1}$ for a certain value of $x_{c}$ in the range $1\leq x_{c}\leq 10$ for the parameter values used in Fig.~\ref{fg:pole}. (2) Even for $x_{c}=1$ it can become smaller than $\kappa_{1}$ as the ratio $U_{\infty}/\kappa_{1}$ increases. For example, we obtain  $|\text{Im}(\omega_{0})|\simeq 0.183$ for $U_{\infty}=40$, $\kappa_{1}=0.5$ and $x_{c}=l=1$. We can claim that the conditions $\text{Re}(\omega_{q})\neq 0$ and $|\text{Im}(\omega_{q})|<\kappa_{1}$ are satisfied in a wide range of the parameter values.


\end{document}